%% file: main.tex
\documentclass{vldb}

\makeatletter
\newif\if@restonecol
\makeatother

\usepackage[lined,boxed,vlined,ruled]{algorithm2e}
\usepackage{balance}
\usepackage{algorithmicx}
\usepackage{hyperref}
\usepackage{verbatim}
\usepackage{pifont}
\usepackage[T1]{fontenc}
\usepackage{textcomp}
\usepackage{caption}
\usepackage{subcaption}
\usepackage[lined,boxed,vlined,ruled]{algorithm2e}
\usepackage{listings}
\usepackage{framed}
\usepackage{xcolor}
\usepackage{geometry}
\usepackage{changepage}
\colorlet{shadecolor}{gray!20}

\input{src/commands}

\vldbTitle{An End-to-End Learning-based Cost Estimator}
\vldbAuthors{Ji Sun, Guoliang Li}
\vldbDOI{https://doi.org/10.14778/xxxxxxx.xxxxxxx}
\vldbVolume{12}
\vldbNumber{xxx}
\vldbYear{2018}

\begin{document}
\title{An End-to-End Learning-based Cost Estimator}
\numberofauthors{1}
\author{
\alignauthor
Ji Sun~~~~~~~Guoliang Li\\
       \affaddr{Department of Computer Science, Tsinghua~University}\\
       \affaddr{sun-j16@mails.tsinghua.edu.cn,liguoliang@tsinghua.edu.cn}
}

\maketitle

\begin{abstract}
Cost and cardinality estimation is vital to query optimizer, which can guide the plan selection. However traditional empirical cost and cardinality estimation techniques cannot provide high-quality estimation, because they cannot capture the correlation between multiple columns. Recently the database community shows that the learning-based cardinality estimation is better than the empirical methods. However, existing learning-based methods have several limitations.  Firstly, they can only estimate the cardinality, but cannot estimate the cost. Secondly, convolutional neural network (CNN) with average pooling is hard to represent complicated structures, e.g., complex predicates, and the model is hard to be generalized. 


To address these challenges, we propose  an effective end-to-end learning-based cost estimation framework based on a tree-structured model, which can estimate both cost and cardinality simultaneously. To the best of our knowledge, this is the first end-to-end cost estimator based on deep learning.  We propose  effective feature extraction and encoding techniques, which consider both queries and physical operations in feature extraction. We embed these features into our tree-structured model. We propose an effective method to encode string values, which can improve the generalization ability for predicate matching. As it is prohibitively expensive to enumerate all string values, we design a patten-based method, which selects patterns to cover string values and utilizes the patterns to embed string values.  We conducted experiments on real-world datasets and experimental results showed that our method outperformed baselines.
\end{abstract}

\input{src/introduction}

\input{src/related-work}

\input{src/frameworks}

\input{src/model}

\input{src/str-pretrain}
\input{src/exp}
\input{src/sec-con}

\newpage
\balance

\input{main.bbl}
\balance

\end{document}

%% file: src/commands.tex
\newcommand{\eat}[1]{}
\usepackage{latexsym}
\usepackage{amsfonts}
\usepackage{amsmath}
\usepackage{amssymb}
\usepackage{color}
\usepackage{colortbl}
\usepackage{epsfig}
\usepackage{xspace}
\usepackage{graphicx}

\usepackage{paralist}
\usepackage{enumerate}
\usepackage{pifont}
\usepackage{ifthen}
\newboolean{isTechReport}
\setboolean{isTechReport}{true}   
\newcolumntype{L}[1]{>{\raggedright\let\newline\\\arraybackslash\hspace{0pt}}m{#1}}
\newcolumntype{C}[1]{>{\centering\let\newline\\\arraybackslash\hspace{0pt}}m{#1}}
\newcolumntype{R}[1]{>{\raggedleft\let\newline\\\arraybackslash\hspace{0pt}}m{#1}}

\newtheorem{definition}{Definition}

\newcommand{\hi}[1]{\vspace{.25em} \noindent {\bf #1}\xspace}
\newcommand{\highlight}[1]{{\vspace{.25em}\noindent{\bf #1}}\xspace}
\newcommand{\tabincell}[2]{\begin{tabular}{@{}#1@{}}#2\end{tabular}}
\newcommand{\pgcard}{\ensuremath{\texttt{PGCard}}\xspace}
\newcommand{\mscncard}{\ensuremath{\texttt{MSCNCard}}\xspace}
\newcommand{\mscnnosamplecard}{\ensuremath{\texttt{MSCNNSCard}}\xspace}
\newcommand{\tnncard}{\ensuremath{\texttt{TNNCard}}\xspace}
\newcommand{\tlstmcard}{\ensuremath{\texttt{TLSTMCard}}\xspace}
\newcommand{\tlstmnosamplecard}{\ensuremath{\texttt{TLSTMNSCard}}\xspace}

\newcommand{\tlstmhashcard}{\ensuremath{\texttt{TLSTMHashCard}}\xspace}

\newcommand{\tlstmembwcard}{\ensuremath{\texttt{TLSTMEmbNRCard}}\xspace}
\newcommand{\tlstmembtcard}{\ensuremath{\texttt{TLSTMEmbRCard}}\xspace}
\newcommand{\tpembtcard}{\ensuremath{\texttt{TPoolEmbRCard}}\xspace}

\newcommand{\pgcost}{\ensuremath{\texttt{PGCost}}\xspace}
\newcommand{\mscncost}{\ensuremath{\texttt{MSCNCost}}\xspace}

\newcommand{\tlstmcostonly}{\ensuremath{\texttt{TLSTMCost}}\xspace}
\newcommand{\tnncost}{\ensuremath{\texttt{TNNMCost}}\xspace}
\newcommand{\tlstmcost}{\ensuremath{\texttt{TLSTMMCost}}\xspace}

\newcommand{\tlstmhashcost}{\ensuremath{\texttt{TLSTMHashMCost}}\xspace}

\newcommand{\tlstmembwcost}{\ensuremath{\texttt{TLSTMEmbNRMCost}}\xspace}
\newcommand{\tlstmembtcost}{\ensuremath{\texttt{TLSTMEmbRMCost}}\xspace}
\newcommand{\tpembtcost}{\ensuremath{\texttt{TPoolEmbRMCost}}\xspace}

\newcommand{\pg}{\ensuremath{\texttt{PostgreSQL}}\xspace}
\newcommand{\mscn}{\ensuremath{\texttt{MSCN}}\xspace}
\newcommand{\mscnb}{\ensuremath{\texttt{MSCNBatch}}\xspace}
\newcommand{\tpe}{\ensuremath{\texttt{TPool}}\xspace}
\newcommand{\tpbe}{\ensuremath{\texttt{TPoolBatch}}\xspace}
\newcommand{\tlstme}{\ensuremath{\texttt{TLSTM}}\xspace}
\newcommand{\tlstmbe}{\ensuremath{\texttt{TLSTMBatch}}\xspace}

\usepackage{epsfig}
\usepackage{multirow}
\usepackage{url}

\definecolor{orange}{HTML}{FF7F00}

\newcommand{\bi}{\begin{itemize}}
\newcommand{\ei}{\end{itemize}}

\newcommand{\be}{\begin{enumerate}}
\newcommand{\ee}{\end{enumerate}}
\newcommand{\beqn}{\begin{eqnarray*}}
\newcommand{\eeqn}{\end{eqnarray*}}

\newcommand{\eg}{{e.g.,}\xspace}

\newcounter{ccc}

\newcommand{\eop}{\hspace*{\fill}\mbox{$\Box$}\vspace{1ex}}     

\newcommand{\nthesection}{\arabic{section}}
\renewcommand{\thetheorem}{\arabic{theorem}}
\newcounter{prop}[section]

\newcounter{cor}
\renewcommand{\thecor}{\arabic{theorem}}

\newcounter{alg}[section]
\renewcommand{\thealg}{\nthesection.\arabic{alg}}

\newcounter{arule}
\renewcommand{\thearule}{\arabic{arule}}

\newcounter{claim}
\renewcommand{\theclaim}{\arabic{claim}}

\makeatletter
    \newcommand\figcaption{\def\@captype{figure}\caption}
    \newcommand\tabcaption{\def\@captype{table}\caption}
\makeatother

\definecolor{shadecolor1}{RGB}{230,230,230}

\definecolor{shadecolor1}{RGB}{255, 114, 118}

\usepackage[all]{nowidow}


%% file: src/introduction.tex

\section{Introduction}

Query optimizer is a vital component of database systems, which aims to select an optimized query plan for a SQL query. However, recent studies show that the classical query optimizer~\cite{DBLP:journals/pvldb/LeisGMBK015,DBLP:conf/cidr/LeisRGK017,optimization_solved} often generates sub-optimal plans due to poor cost and cardinality estimation. First, traditional empirical cost/cardinality estimation techniques cannot capture the correlation between multiple columns, especially for a large number of tables and columns. Second, the cost model requires to be fine-tuned by DBAs.

Recently, the database community attempts to utilize machine learning models to improve cardinality estimation. MSCN~\cite{DBLP:conf/cidr/KipfKRLBK19} adopts the convolutional neural network to estimate the cardinality. However, this method has three limitations. Firstly, it can only estimate the cardinality, but cannot estimate the cost. Secondly, the deep neural network with average pooling is hard to represent complicated structures, e.g., complex predicates and tree-structured query plan, and thus the model is hard to be generalized to support most of SQL queries. Thirdly, although MSCN outperforms PostgreSQL in cardinality estimation, its performance can be further improved.  For example, on the JOB-LIGHT workload, the mean error is over 50 and the max error is over 1,000. We can improve them to 24.9 and 289 respectively.

There are four challenges to design an effective learning-based cost estimator. First, it requires to design an end-to-end model to estimate both cost and cardinality. Second, the learning model should capture the tree-structured information of the query plan, e.g., estimating the cost of a plan based on its sub-plans. Third, it is rather hard to support predicates with string values, e.g., predicate "not LIKE `\%(co-production)\%'", if we don't know which values contain pattern `(co-production)'. As the string values are too sparse, it is rather hard to embed the string values into the model. Fourth, the model should have a strong generalization ability to support various of SQL queries.


To address these challenges, we propose an end-to-end learning-based cost estimation framework by using deep neural network. We design a tree-structured model that can learn the representation of each sub-plan effectively and can replace traditional cost estimator seamlessly. The tree-structured model can also represent complex predicates with both numeric values and string values.

In summary, we make the following contributions.

\noindent(1) We develop an effective end-to-end learning-based cost estimation framework based on a tree-structured model, which can estimate both cost and cardinality simultaneously. To the best of our knowledge, this is the first end-to-end cost estimator based on deep learning (see Section~\ref{sec:overview}).

\noindent(2) We propose effective feature extraction and encoding techniques, which consider  both queries and physical execution in feature extraction. We embed these features into our tree-structured model, which can estimate the cost and cardinality utilizing the tree structure (see Section~\ref{sec:model}).

\noindent(3)  For predicates with string values, we propose an effective method to encode string values for improving the generalization ability. As it is prohibitively expensive to enumerate all possible string values, we design a pattern-based method, which selects patterns to cover string values and utilizes the patterns to embed the string values (see Section~\ref{sec:embedding}).

\noindent(4) We conducted experiments on real-world datasets, and experimental results showed that our method outperformed existing approaches (see Section~\ref{sec:exp}).

%% file: src/related-work.tex
\vspace{-.5em}
 \section{Related Work}

\highlight {\bf Traditional Cardinality Estimation.}
Traditional cardinality estimation techniques can be broadly classified into three classes. The first is \textit{histogram-based} methods~\cite{DBLP:conf/vldb/Ioannidis03}. The core idea is to divide the cell values into equal depth or equal width buckets, keep the cardinality of each bucket, and estimate the cardinality according to the buckets. The method is easy to implement and has been widely used in commercialized databases. However, it is not effective to estimate the correlations between different columns. The second is \textit{sketching}, which aims to solve distinct cardinality estimation problem, including FM~\cite{FLAJOLET1985182}, MinCount~\cite{GIROIRE2009406}, LinearCount~\cite{DBLP:journals/tods/WhangVT90}, LogLog~\cite{10.1007/978-3-540-39658-1_55}, HyperLogLog~\cite{Flajolet07hyperloglog:the}. The basic idea first maps the tuple values to bitmaps, then counts the continuous zeros or the number of hitting for each position, and finally infers the approximate number of distinct values. These methods can estimate the distinct number of rows for each dataset effectively. However, they are not suitable for estimating range query. The third is \textit{sampling-based} methods~\cite{Lipton:1990:PSE:93605.93611,10.1007/3-540-52342-1_23,Wu:2016:SQR:2882903.2882914,DBLP:conf/cidr/LeisRGK017}. These methods utilize the data samples to estimate the cardinality. In order to address the sample vanishing problem (valid samples decrease rapidly for joins), \cite{DBLP:conf/cidr/LeisRGK017} proposed index-based sampling. Sampling methods improve the accuracy of cardinality estimation, but they bring space overhead and only be adopted by in-memory database like HyPer~\cite{DBLP:conf/btw/0001LK17}. Another limitation of this method is 0-tuple problem, i.e., when a query is sparse, if the bitmap equals to 0, the sample is invalid.

\highlight {\bf Traditional Cost Model.}
Traditional cost estimation is estimated by combining multiple factors like cost of sequential page fetch, cost of random page fetch, cost of CPU cost of processing a tuple and cost of performing operation. Firstly,  these factors are highly correlated to the cardinality of data affected by the query. Secondly, the weight of each factor has to be tuned. There are some works focusing the cost model tuning~\cite{6544899,Liu:2015:FCP:2806777.2806944,DBLP:journals/pvldb/LeisGMBK015}, and \cite{DBLP:journals/pvldb/LeisGMBK015} conducted experiments on the IMDB dataset to show that cardinality estimation is much more crucial than the cost model for cost estimation and query optimization.

\highlight {\bf Learning-based Cardinality Estimation.}
The database community starts to solve this  problem by using learning-based method like statistic machine learning or deep neural network. The first learning based work on cardinality estimation~\cite{DBLP:conf/cidr/MalikBC07} first classifies queries according to the query structure (join condition, attributes in predicates etc.), and then trains a model on the  values of the predicates, but the model is ineffective to train on unknown structured query. The state-of-the-art method~\cite{DBLP:conf/cidr/KipfKRLBK19} trains a multi-set convolutional network on queries, but this method is not suitable for query optimization, because the query-based encoding is too tricky when optimizing on a tree structure, and the generalization is limited. \cite{DBLP:conf/sigmod/OrtizBGK18} proposed a vision of training representation for the join tree with reinforcement learning. However, this method does not support query plans with a complex tree structure and cannot support complex queries.

\highlight {\bf Learning-based Performance Prediction.} There are several works on performance prediction by using machine learning and statistics~\cite{Akdere:2012:LQP:2310257.2310339,6544899,Li:2012:RER:2350229.2350269,Ganapathi:2009:PMM:1546683.1547490,Zhang:2005:SLT:1083592.1083628}, but they  all require experts to select the features according to operation properties. A deep learning based approach~\cite{DBLP:journals/corr/abs-1902-00132} is also proposed. However, this method must take the estimated cardinality and the cost of PostgreSQL as features without learning the semantic of predicates, and it is not an end-to-end solution.

%% file: src/frameworks.tex

\vspace{-.5em}

 \section{Overview of End-to-end Cost Estimator}
\label{sec:overview}

Cost estimation is to estimate the execution cost of a query plan, and the estimated cost is used by the query optimizer to select physical plans with low cost. Cardinality estimation is to estimate the number of tuples in the result of a (sub)query.  In this section we propose the system overview of  cost  and  cardinality estimation.

\begin{figure}[!t]
\begin{center}
\centering
\includegraphics[width=0.35\textwidth]{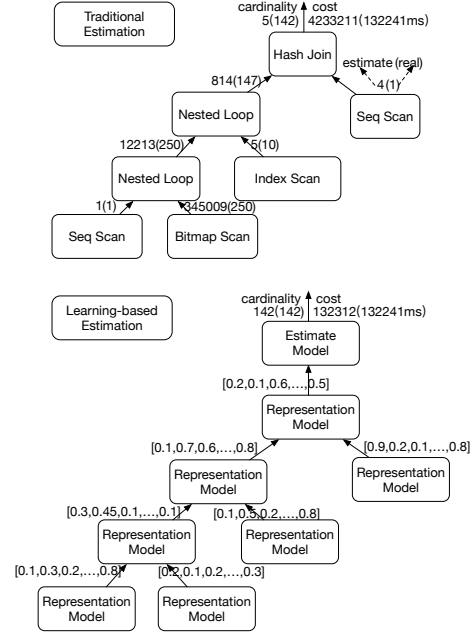}
\vspace{-.5em}
\caption{Comparison of Traditional Cost Estimation and Learning-based Cost  Estimation. \label{fig:replace}}
\end{center}
\vspace{-2em}
\end{figure}

\begin{figure*}[!t]
\begin{center}
\centering
\includegraphics[width=1.0\textwidth]{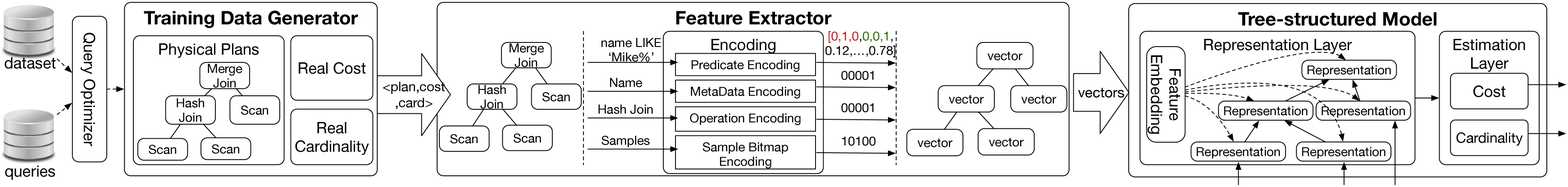}
\vspace{-2em}
\caption{Architecture of learning-based cost estimator\label{fig:framework}}
\end{center}
\vspace{-3em}
\end{figure*}

\begin{figure*}[t]\vspace{1em}
\begin{center}
\centering
\includegraphics[width=1.0\textwidth]{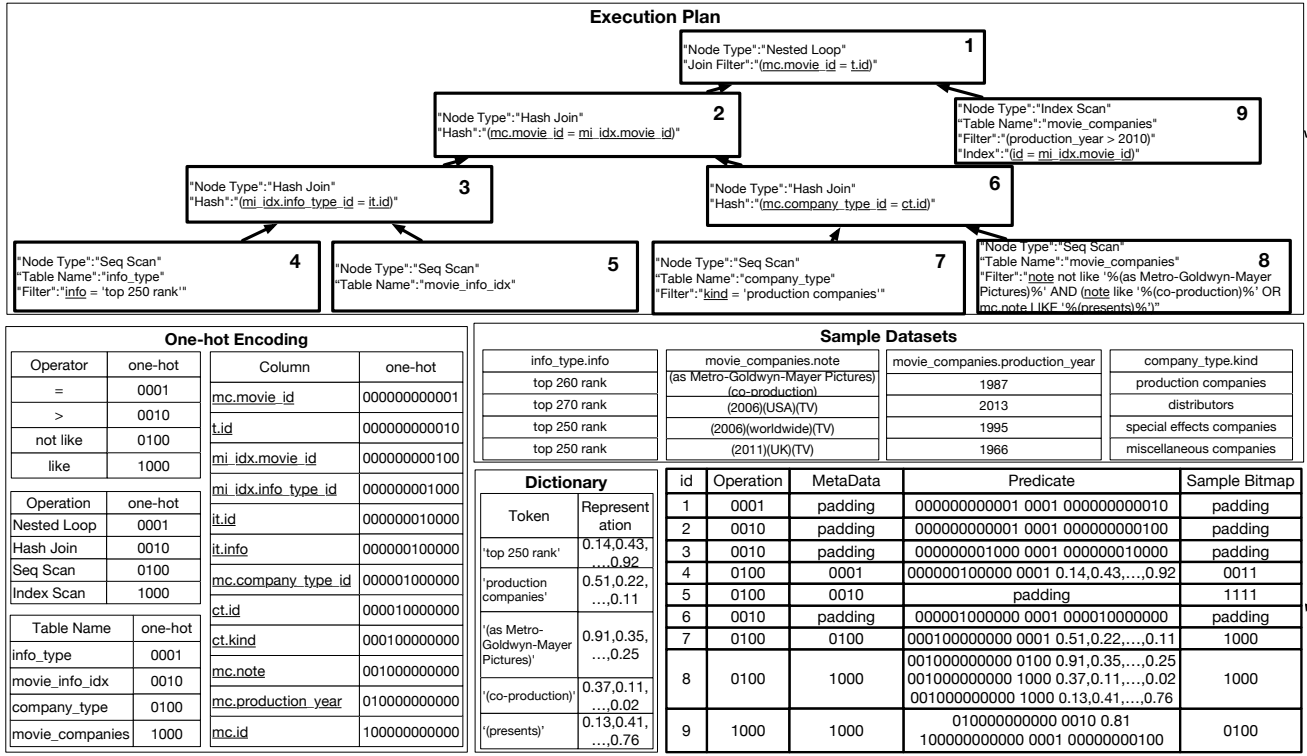}
\vspace{-2em}
\caption{Running Example of query plan encoding\label{fig:running-example}}
\end{center}
\vspace{-2.5em}
\end{figure*}

Traditional databases estimate the cost and cardinality using statistics. For filter operations, cardinality estimator (e.g., PostgreSQL\cite{pg}, DB2\cite{ibm_db2}) estimates the cardinality using the  histograms; for join operations, the cardinality is estimated by empirical functions with selectivity of joined tables (nodes) as variables. In Figure~\ref{fig:replace}, the numbers on top of each node are estimated cardinality and real cardinality. We    find that there exist large errors in traditional methods.

In general, we can effectively estimate the cardinality for leaf nodes (like Scan) by using the histogram; however, the error would be very large for joins because of the correlations between tables. Usually the more joins are, the larger error is. Unlike traditional cost estimation methods, our learning-based model can learn the correlation among multiple columns and tables, and the representation can retain accurate information on distribution of results even for the queries with dozens of operations.

Moreover, the query plan is a tree structure, and the plan is executed in a bottom-up manner.  Intuitively, the cost/cardinality of a plan should be estimated based on its sub-plans. To this end, we design a tree-structured model that matches the plan naturally, where each model can be composed of some sub-models in the same way as a plan is made up of sub-plans.  We use the tree-structured model to estimate  the cost/cardinality of a plan in a bottom-up manner. 

\hi{Learning-based Cost Estimator.} The end-to-end learning-based tree-structured cost estimator includes three main components, including training data generator, feature extractor, and tree-structured model, as shown in Figure~\ref{fig:framework}.

\noindent
1) \textbf{Training Data Generator} generates training data based on the data and query workload.  It first generates a large number of queries according to the potential join graph of the dataset and the predicates in the workload. Then for each query, it extracts a physical plan by the optimizer and gets the real cost/cardinality. Thus a training data is a triple $\langle$a physical plan, the real cost of the plan, the real cardinality of the plan$\rangle$.

\noindent
2) \textbf{Feature Extractor} extracts useful features from the query plan, e.g.,  query operation and predicates. Each node in the query plan is encoded into feature vectors and each vector is organized into tensors. Then the tree-structured vectors are taken as input of the training model. For simple features, we can encode them by using one-hot vector or bitmap. While for complicated features, e.g., LIKE predicate, we encode each tuple $\langle$column, operator, operand$\rangle$ into vectors, by using a one-to-one mapping (see Section~\ref{sec:model:feature}).

\noindent
3) \textbf{Tree-structured Model} defines a tree-structured model which can learn representations for the (sub)plans, and the representations can be used in cost and cardinality estimation. The model is trained based on the training data, stores the updated parameters in the model, and estimates cost and cardinality for new query plans.

\hi{\bf Workflow.} For offline training, the training data are generated by \textit{Training Data Generator}, which are encoded into tensors by \textit{Feature Extractor}. Then the training data is fed into the \textit{Training Model} and the model updates weights by back-propagating based on current training loss. The details of model training is discussed in Section~\ref{sec:model:train}.

For online cost estimation, when the query optimizer asks the cost of a  plan, \textit{Feature Extractor} encodes it in a up-down manner recursively. If the sub-plan rooted at the current node has been evaluated before, it extracts representation from \textit{Representation Memory Pool}, which stores a mapping from a  query plan to its estimated cost.  If the current sub-plan is new, {\it Feature Extractor} encodes the root and goes to its children nodes. We input the encoded plan vector into \textit{Tree-structured Model}, and then the model evaluates the cost and cardinality of the plan and returns them to the query optimizer. Finally, the estimator puts all the representations of `new' sub-plans into  \textit{Representation Memory Pool}.

Figure~\ref{fig:running-example} shows a running example of feature encoding for a plan extracted from the JOB workload. The plan is encoded as a vector using the one-hot encoding scheme, which considers both query and database samples. Then the vectors are taken as an input of the training model.

%% file: src/model.tex
\vspace{-.5em}
\section{Tree-based Learning Model}
\label{sec:model}

In this section, we introduce a tree-structured deep neural network based solution for end-to-end cost estimation. We first introduce feature extraction in Section~\ref{sec:model:feature}, and then discuss model design in Section~\ref{sec:model:model}. Finally we present the model training in Section~\ref{sec:model:train}.

\vspace{-.5em}
\subsection{Feature extraction and encoding}
\label{sec:model:feature}

We first encode a query node as a node vector, and then transform the node vectors into a tree-structured vector.   

There are four main factors that may affect the query cost, including the physical query operation, query predicate,  meta data,  and data.  Next we discuss how to extract these features and encode them into vectors.

\hi{Operation} is the physical operation used in the query node, including {\it Join} operation (e.g., Hash Join, Merge Join, Nested Loop Join); {\it Scan} operation (e.g., Sequential Scan, Bitmap Heap Scan, Index Scan, Bitmap Index Scan, Index Only Scan); {\it Sort} operation (e.g., Hash Sort, Merge Sort); {\it Aggregation} operation (e.g., Plain Aggregation, Hash Aggregation). These operations significantly affect the cost.  Each operation can be encode as a one-hot vector.  Figure~\ref{fig:running-example} shows the one-hot vectors of different operations.

\hi{Predicate} is the set of filter/join conditions used in a node. The predicate may contain join conditions like `movie.movie\_id = mi\_idx.movie\_id' or filter conditions like `production\_year > 1988'. Besides the {\it atomic predicates} with only one condition, there may exist {\it compound predicates} with multiple conditions, like `production\_year > 1988 AND production\_year < 1993'. The predicates affect the query cost, because the qualified tuples will change after applying the predicates.

\begin{table}[!t]
\center
\caption{Main Plan Operations\label{table:plan:feature}}
\vspace{-1em}
\begin{tabular}{|c|c|}
\hline
Operation & Features \\ \hline
$Aggregate$ & $[Operator,Name_{keys}]$ \\\hline
$Sort$ & $[Operator,Name_{keys}]$ \\\hline
$Join$ & $[Operator,Predicate_{join}]$ \\\hline
$Scan$ & \tabincell{c}{$[Operator,Name_{table},Name_{index},$\\$Predicate_{filter}, SampleBitmap]$} \\\hline
\end{tabular}
\vspace{-1.5em}
\center
\end{table}

\begin{table}[!t]\vspace{1em}
\center
\vspace{-1.5em}
\caption{Features of Condition Operators \label{table:condition:feature}}
\vspace{-1em}
\begin{tabular}{|c|c|c|}
\hline
Operators & Features \\ \hline
and/or/not & $[Operator]$ \\\hline
=/!=/>/</LIKE/IN & \tabincell{c}{$[Operator,Column,$ $Operand]$} \\
\hline
\end{tabular}
\vspace{-2.5em}
\center
\end{table}

Each atomic predicate is composed of three parts, {\it Column}, {\it Operator}, and {\it Operand}. {\it Operator} and {\it Column} can be encoded as one-hot vectors. For {\it Operand}, if it is a numeric value, we encode it by a normalized float; if it is string value, we encode it with a string representation (see Section~\ref{sec:embedding} for details). Then the vector of an atomic predicate is the concatenation of the vectors for column, operator and operand. Table~\ref{table:condition:feature} shows the vector of each predicate.

\begin{figure*}[!t]\vspace{1em}
\begin{center}
\centering
\includegraphics[width=1\textwidth]{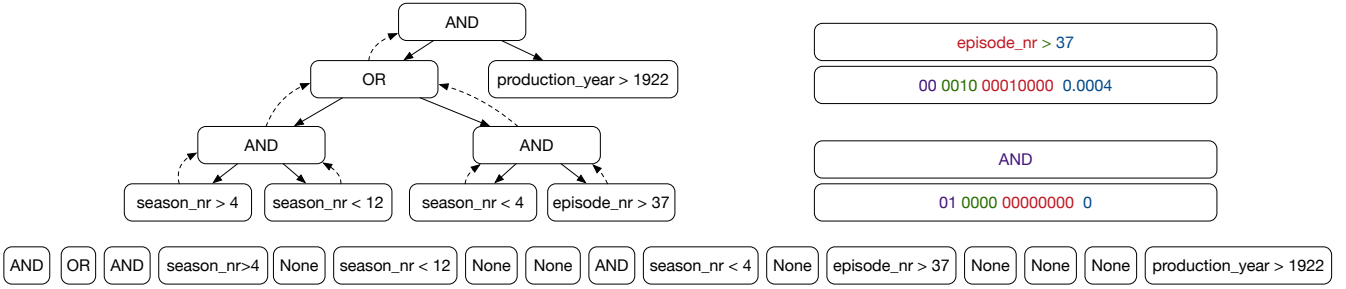}
\caption{Predicates Encoding\label{fig:tree_coding}}
\end{center}
\vspace{-1.5em}
\end{figure*}

For a compound predicate, we first generate a vector for each atomic predicate and then transfer multiple vectors into a vector using a one-to-one mapping strategy. There are multiple ways to transfer a tree structure to a sequence in a one-to-one mapping, and here we take the depth first search (DFS) as an example. We first transfer the nodes to a sequence using DFS, and then concatenate the vectors  following the sequence order.  Figure~\ref{fig:tree_coding} shows an example of encoding a compound predicate.  We transfer the nodes into a sequence in the visited order, where the solid lines with arrow represent forward search, and each dotted line represents one step backtracking. We append an empty node to the end of sequence for each dotted line. Thus, we can encode each distinct complex predicate tree as a unique vector and  the compound predicate can be encoded as a tensor.

\hi{MetaData} is the set of columns, tables and indexes used in a query node. We use a one-hot vector for columns, tables, and indexes respectively. Then the meta node vector is a concatenation of column vectors, table vectors and table vectors. (Note that a node may contain multiple tables, columns and indexes, and we can compute the unions of different vectors using the OR semantic.) We encode both meta data and predicate, because some nodes may not contain predicates.  Figure~\ref{fig:tree_coding} shows an example.

\hi{Sample Bitmap} is a fix-sized 0-1 vector where each bit denotes whether the corresponding tuple satisfies the predicate of the query node. If the data tuple matches the predicate, the corresponding bit is 1; 0 otherwise. This feature is only included in nodes with predicates. Since it is expensive to maintain a vector for all tuples, we select some samples for each table and maintain a vector for the samples.  Figure~\ref{fig:tree_coding} shows an example of encoding sample data.

After encoding each node in the plan, we need to encode the tree-structured plan into a vector using a one-to-one mapping strategy. We also adopt the DFS method in the same ways as encoding compound predicates.

\subsection{Model Design}
\label{sec:model:model}

\begin{figure*}[!t]\vspace{1em}
\begin{center}
\centering
\includegraphics[width=1\textwidth]{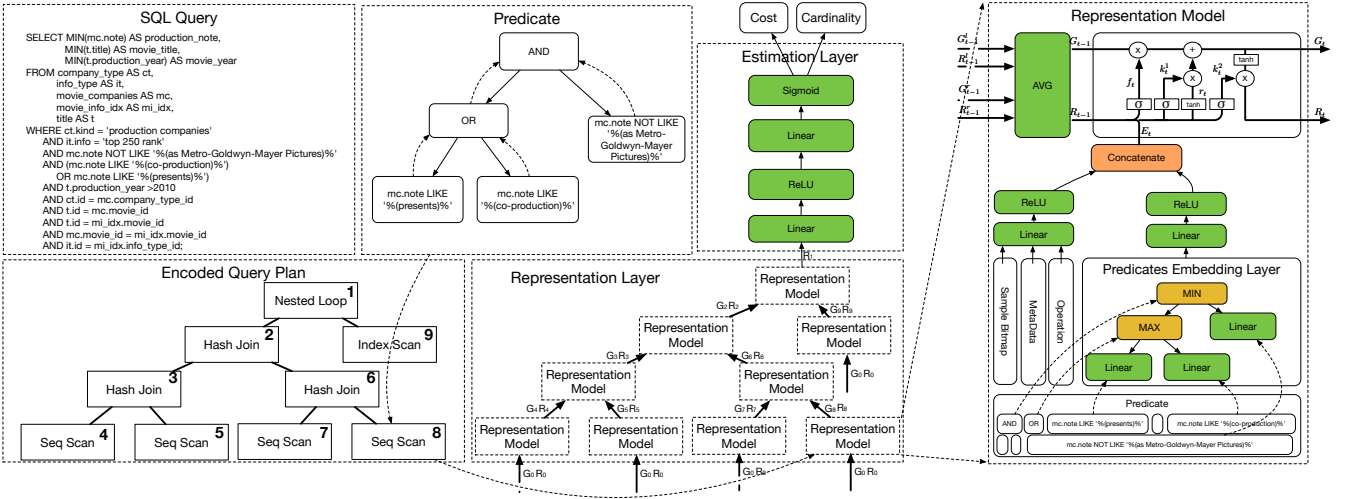}
\caption{Two Level Tree Model\label{fig:model}}
\end{center}
\vspace{-2.5em}
\end{figure*}

Our model is composed of three layers, \textit{embedding layer}, \textit{representation layer} and \textit{estimation layer} as shown in Figure~\ref{fig:model}. Firstly,  feature vectors in each plan node are large and sparse, we should condense them and extract high-dimensional information of features, and thus the embedding layer embeds the vector for each plan node. Secondly, the {\it representation layer} employs a tree-structured model, where each node is a representation model and the tree structure is the same as the plan. Each representation model learns two vectors (global vector and local vector) for the corresponding sub-plan, where the global vector captures the information of the sub-plan rooted at the node and the local vector captures  the information of the node. Node that each representation model learns the two vectors based on the vectors of its two children and the feature vector of the corresponding node.  Finally, based on the vectors of the root node, \textit{estimation layer}  estimates the cost and cardinality. 

\vspace{-.5em}
\subsubsection{Embedding Layer}
\label{sec:model:embedding}

The embedding layer aims to embed a sparse vector to a dense vector. 

As discussed in Section~\ref{sec:model:feature}, there are 4 types of features, {\it Operation}, {\it Metadata},  {\it Predicate} and {\it Sample Bitmap}. {\it Operation} is a one-hot encoding vector, and {\it Metadata} and {\it Sample Bitmap} are bitmap vectors. We use a one-layer fully connected neural network with ReLU activator to embed these three vectors. However, the structure of the {\it Predicate} vector is complicated because it contains multiple AND/OR semantics, and we design an effective  model to learn the representation of predicates. 

Our goal is to estimate the number of tuples that satisfy a predicate. For an atomic predicate, we can directly use the vector. But for a compound predicate with multiple conditions, we need to learn the semantic of the predicates and the distribution of the results after applying the predicates on the dataset. 

Consider a compound predicate with two atomic predicates using the AND semantic. We can estimate the number of results satisfying the predicate by the minimum number of estimated results satisfying the atomic predicates. Thus we use the \textit{min} pooling layer to combine the two atomic predicates.  
Consider a compound predicate with two atomic predicates using the OR semantic. We can estimate the  number of results satisfying the predicate by the maximum number of estimated results satisfying the atomic predicates. Thus we use the \textit{max} pooling layer to combine the two atomic predicates.  

In this way, we use a tree pooling to encode a predicate, where the tree structure is the same as the predicate tree structure. Particularly, the leaf node is a fully connected neural network, the OR semantic is replaced with the max pooling layer and the AND semantic is replaced with the min pooling layer.  The advantages of this model are two folds. The first is that only the leaf nodes need to be trained so that it's easy to do efficient batch training. The second is that this model converges faster and performs better.

Figure~\ref{fig:tree_coding} shows a compound predicate and its embedded model.  For leaf nodes, we use a fully connected neural network. For conjunction nodes, we use max pooling layer for `OR' and min pooling layer for `AND' which meet the semantic of `AND' and `OR'.

\hi{Embedding Formulation.} 
We denote features {\it Operation}, {\it Metadata},  {\it Predicate} and {\it Sample Bitmap} of $node_t$ as $O_t, M_t, P_t, B_t$ respectively, and we denote each node of feature {\it Predicate} as $P_t$ with $P^l_t$ as its left child and $P^r_t$ right child. {\it Embedding Model} can be formalized as below. $E$ is the embedding output. $W$ is the weight of a fully connected neural network. $b$ is a bias. 
\begin{equation*}
\begin{split}
&E =  [embed(O_t), embed(M_t), embed(B_t), embed(P_t)] \\ 
&embed(O_t) =  ReLU(W_{o} \cdot O_t + b_{o}) \\
&embed(M_t) =  ReLU(W_{m} \cdot M_t + b_{m}) \\
&embed(B_t) =  ReLU(W_{b} \cdot B_t + b_{b}) \\
&embed(P_t) = \left\{
             \begin{array}{lr}
             min(embed(P^l_t), embed(P^r_t)) & type(P_t)=AND, \\
             max(embed(P^l_t), embed(P^r_t)) & type(P_t)=OR,\\
             W_{p} \cdot P_t + b_{p} & type(P_t)=Exp.
             \end{array}
\right.\\
\end{split}
\end{equation*} 
where $type(P_t)$ is the type of a node, which includes AND, OR, and a predicate expression.

\subsubsection{Representation Layer}
\label{sec:model:represent}

Cost estimation has two main challenges -- information vanishing and space explosion. First,  it is easy to estimate the cost for simple operations, e.g., estimating the cost of a filtering predicate on a single table, but it is rather hard to estimate the cost of joining multiple tables, because the join space is large and the joined tuples are sparse. In other words, in leaf nodes, we can capture much information for single table processing. But for upper nodes in the query plan, the correlation among nodes may be lost and this is the information vanishing problem. Second, to retain enough information to capture the correlations among tables, it requires to store much more space and intermediate results. But the space grows exponentially and becomes prohibitively expensive for a large number of tables.  This is the space explosion problem. 


Representation layer aims to address these two problems, by capturing the global cost information from leaf nodes to the root and avoiding information loss (e.g., correlation between columns). 
The representation layer trains representation for sub-plan recursively, instead of using data sketch to represent intermediate results of sub-plans. The representation layer uses the representation vector learned from features to represent results of sub-plans. As Figure~\ref{fig:model} shows, all the units in this layer are neural networks in the same structure and share common parameters and we call these units as representation models. Each representation model has three inputs, the {\it embedding vector} $E$, the representation vector $[G^l_{t-1},R^l_{t-1}]$ of its left child,  and the representation vector $[G^r_{t-1},R^r_{t-1}]$ of its right child. (Note that for leaf nodes, we use zero vectors as their children vectors.)  The output is  $[G_{t},R_{t}]$.

The most important design issue in this layer is the choice of recurrent neural network. As a joint network of the tree-structured model, it decides which information to be passed over. A naive implementation is using fully connected neural network which takes as input the concatenate of local transformed features and its  children's output. However, the lost information would never be utilized by upper nodes any more. Therefore, representation model with naive neural networks suffers from gradient vanishing and gradient explosion problems.

Compared to naive deep neural network, LSTM cell can efficiently address these problems by using an extra information channel. The structure of LSTM cell is shown in Figure \ref{fig:model}, where $G_t$ is the channel for long memory, $f_t$ controls which information should be forgot in the long memory. $k^1_t$ controls which information should be added into the long memory channel. $k^2_t$ controls which information in the memory channel should be taken as the representation of the sub-plans. As  $G_t$ can be a path without any multiplication, LSTM avoids gradient vanishing. The forget gate of Sigmoid can help LSTM to address gradient explosion problem. 
The representation layer can be formalized as below:

\vspace{-1.5em}
\begin{equation*}
\begin{split}
&x_t = E \\
&G_{t-1} = (G^l_{t-1}+G^r_{t-1}) / 2 \\
&R_{t-1} = (R^r_{t-1}+R^r_{t-1}) / 2 \\
&f_t = Sigmoid(W_f \cdot [R_{t-1},x_t]+b_f) \\
&k^1_t = Sigmoid(W_{k_1} \cdot [R_{t-1},x_t]+b_{k_1}) \\
&r_t = tanh(W_r \cdot [R_{t-1},x_t]+b_r) \\
&k^2_t = Sigmoid(W_{k_2} \cdot [R_{t-1},x_t]+b_{k_2}) \\
&G_t = f_t \times G_{t-1} + k^1_t \times r_t\\
&R_t = k^2_t \times tanh(G_t) \\
\end{split}
\end{equation*}
\vspace{-.5em}
where $E$ is the vector, $W$ is a weight and $b$ is a bias. 

\subsubsection{Estimation Layer}
\label{sec:model:estimate}

The {\it Estimation Layer} is a two-layer fully connected neural network, and the activator is a ReLU function and the output layer is a sigmoid function to predict the normalized cardinality and cost. The output layer should be able to evaluate cost or cardinality for any sub-plan by its representation vector.  This layer takes the representation $R_t$ of the upper model in {\it Representation Layer} as input, and can be formalized as below:
\vspace{-.5em}
\begin{equation*}
\begin{split}
&cost' = ReLU(W_{cost'} \cdot R_t + b_{cost'}) \\
&card' = ReLU(W_{card'} \cdot R_t + b_{card'}) \\
&cost = Sigmoid(W_{cost} \cdot cost' + b_{cost}) \\ 
&card = Sigmoid(W_{card} \cdot card' + b_{card}) \\
\end{split}
\end{equation*}

As we need to estimate both cost and cardinality,  we use multitask learning~\cite{DBLP:journals/ml/Caruana97} in order to improve the generalization by using the domain information contained in the training signals of related tasks as an inductive bias. Parameter sharing is one of the most common implementation strategies for multitask learning, which means that we train several tasks simultaneously, and the models of these tasks share some network layers. 

Since the cost of the query plan is correlated to the cardinality of each node in the plan tightly, we train the cost estimation task and cardinality estimation task simultaneously. These two tasks share common {\it embedding layer} and {\it representation layer} and we expect these two layers to learn general representations for both tasks, and the {\it estimation layer} can extract cost and cardinality from representations respectively. In this way, the generality of the learning model is better than training  on cost only task, and thus the model can achieve much better performance.


\subsection{Training Model}
\label{sec:model:train}

\hi{Training Data Generation.} Our model has strong generalization ability. Given a dataset and a workload, e.g., IMDB, we generate  potential join relations from the PK-FK index and query workload. According to the join graph, we select the number of tables ($N$) involved in the query workload, and then we select $N$ connected tables. For predicates, we generate numeric expressions and string expressions individually. For numeric expressions, we randomly select $M$ columns with numeric type in the chosen tables and select a value from dataset for each column. Then we pick operators from `$>,<,=,!=$' for each column. While for expressions with string value, we first pick operators from `$=,!=,LIKE,NOT~ LIKE,IN$', and then select values from substring dictionary (see Section~\ref{sec:embedding}). After generating expressions for each table, we aggregate these expressions into complex predicates by using the `AND'/`OR' semantics. For projection, we select several columns in tables as outputs, and select MIN, MAX, COUNT or Non for each attribute. After we obtain all the training SQL queries, we get physical plans from DBMS by using plan analysis tools. Then we transform plans to sequences and divide them into mini-batches to train.

\hi{Loss Function.} Our model trains cost and cardinality simultaneously. The loss function can be a linear combination of cost loss and cardinality loss, and the weight can be regarded as hyper-parameters. We try different loss weights in $\{0.1,0.2,0.5,1,2,5,10\}$, and pick the one with the lowest validation error by cross validation. In order to achieve high quality and accelerate convergence speed, we take the normalized true cardinality/cost as the targets. The loss functions are formalized as below:
\vspace{-.5em}
\begin{equation*}
\begin{split}
loss = \frac{1}{n} \sum_i \bigl(\omega \cdot qerror(cost_i,estimated\_cost_i) +\\
 qerror(card_i, estimated\_card_i)\bigr) 
\end{split}
\end{equation*}
where $n$ is the number of training queries,  $cost_i$ and $card_i$ are respectively the real cost and cardinality, and $estimated\_cost_i$ and $estimated\_card_i$ are respectively the estimated cost and cardinality.
$$qerror(cost_i,estimated\_cost_i) =  \frac{\max(cost_i,estimated\_cost_i)}{\min(cost_i,estimated\_cost_i)}$$
$$qerror(card_i,estimated\_card_i) =  \frac{\max(card_i,estimated\_card_i)}{\min(card_i,estimated\_card_i)}$$


\hi{Batch Training.} In a batch of training samples with size $\mathcal{N}$, let $\mathcal{D}$ denote the maximum depth of the tree structure. We change the depth-first encoding into width-first encoding, where the input of each batch is organized hierarchically which means the first dimension of the nodes tensor equals to $\mathcal{D}$ while not $\mathcal{N}$. An extra tensor is needed to represent the edge of different layers of trees where each element indicates positions of the left child and right child. For each time, the model batches nodes in certain level of encoded plan trees and trains them all at once. In this way, the model only needs to run LSTM cell for each level of trees ($\mathcal{D}$ times) instead of running for each node, which reduces the training and evaluating time by $\mathcal{O}(\frac{2^\mathcal{D}\mathcal{N}}{\mathcal{D}})$ times.

%% file: src/str-pretrain.tex
 \section{String Embedding}
\label{sec:embedding}

The distribution of numeric values can be learned, even though some values do not appear in the training data, because most numeric values are in a continuous space. For example, predicate $production\_year \in [1990,2000)$ can be inferred from  $production\_year \in [1990,1995)$ and $production\_year \in [1995,2000)$. However, string values are sparse and discrete, and thus hard to learn. So it is much harder to learn the predicates with string values than predicates with numeric values.

There are four intuitive ways to represent a predicate with string values, including \textit{one-hot}, \textit{selectivity}, \textit{sample bitmap},  \textit{hash bitmap}. The \textit{one-hot} embedding maps a string to a bit in the vector. However it cannot estimate an approximate result for unseen string values. The \textit{selectivity} embedding first translates the selectivity of a string value into a numerical value, and then utilizes to the numeric value to embed the string. However, it can not reflect the details on which tuples satisfy the predicate. The \textit{sample bitmap} embedding uses samples to embed a string value, i.e., the string value is 1 if the sample contains it; 0 otherwise. However, it suffers from the 0-tuple problem for sparse predicates.  The {\it hash bitmap} embedding first initializes a zero vector $\mathcal{V}$, and then for each character in the string, calculates its hash value $\mathcal{H}$ and set position $\mathcal{H}\%|\mathcal{V}|$ in the bitmap as 1. The number of character types is small (\eg only 60 distinct characters in the JOB workload, including numbers, characters, punctuations, and other special characters). Therefore, the embedding vectors with hundreds of bits could effectively avoid hash collision even for unseen strings. The {hash-bitmap} embedding carries the characters contained in a string, so we can know the approximate overlap of two strings by taking `AND' on their hash bitmap. However, the {hash bitmap} embedding can only capture the similarity between two strings but  cannot reveal the co-occurrences of two strings.

\hi{Embedding-based Method.} In order to represent the distribution of the string values in a dataset, we need to learn representations for string values in the predicates, and the representations can capture the co-occurrences of strings that co-exist in the same tuple. There are two kinds of possible string predicates in a query. The first is exact matching predicate (\eg using keywords `=' and `IN'), and query strings in exact matching predicates will match the tuple values. The second is pattern matching predicate (\eg using keywords `LIKE'), and query strings in pattern matching predicates match substrings of some values in the dataset. In this section, our goal is to pre-train all the substrings in the predicates so that each substring can be encoded by coexistence-aware representation. There are two {challenges}. The first is how to build a dictionary covering all the strings/substrings in both current and future workloads and the size of the dictionary is bounded. The second is how to maintain all the substrings in the dictionary to efficiently  get the encoding of each string/substring with little space overhead. We first give the overview of the embedding method in Section~\ref{subsec:stringembedding}, and then address these two challenges in Sections~\ref{sec:string:extract} and \ref{sec:string:structure} respectively.

\vspace{-.5em}
\subsection{String Embedding Overview} \label{subsec:stringembedding}

Given a dataset and a query workload, we aim to encode all the strings/substrings that either are used in the current workload or will be  used in the future workload. Thus we do not just encode the plain strings appearing in the query workload. Instead, we generalize the strings/substrings and generate some important rules which could extract all the strings/substrings in the query workload from the dataset. Then we extract all substrings that satisfy the rules and train a model to learn a representation for each string/substring. We take a collection of (sub)strings with the key values in one tuple as a sentence and use the skip-gram model~\cite{DBLP:conf/nips/MikolovSCCD13} (a kind of model for word2vec) to train the string embedding. To efficiently get the embedding of each string in order to encode the queries online, we build an index with small space overhead and support fast prefix/suffix searching. 
Thus we address two challenges in string embedding.

\hi{Rule Generation.} We find rules to generalize the (sub)strings in the query workload.

\hi{String Indexing.} We use the patterns to extract all the substrings. Then we encode these substrings. To efficiently get the code of each substring, we construct trie indexes for storing all these substrings with their codes.

\begin{table}[!t]\vspace{1em}
\center
\caption{Example of substring extraction \label{table:rule-gen}}
\vspace{-1em}
\begin{tabular}{|c|c|c|c|}
\hline
Predicate & Extraction  \\ \hline
title LIKE `Din\%' & "Dinos in Kas" $\rightarrow$ "Din" \\ \hline
title LIKE `Sch\%' & "Schla in Tra" $\rightarrow$ "Sch" \\ \hline
title LIKE `\%06\%' & "(2002-06-29)" $\rightarrow$ "06" \\ \hline
title LIKE `\%08\%' & "(2014-08-26)" $\rightarrow$ "08" \\ \hline
\end{tabular}
\vspace{-2em}
\center
\end{table}

\begin{table}[!t]\vspace{1em}
\setlength{\tabcolsep}{1pt}
\small
\center
\caption{Candidate rules for "Dinos in Kas" $\rightarrow$ "Din\%" \label{table:rules:dk}}
\vspace{-1em}
\begin{tabular}{|c|c|c|c|}
\hline
& Rules  \\ \hline
"Din" $\rightarrow$ "Din" & \tabincell{c}{$\langle Prefix, P_t(``Din"), 3 \rangle$ \\ $\langle Prefix, P_CP_t(``in"), 3 \rangle$} \\ \hline
"Dinos" $\rightarrow$ "Din" & \tabincell{c}{$\langle Prefix, P_t(``D")P_l, 3 \rangle$ \\ $\langle Prefix, P_CP_l, 3 \rangle$ \\ $\langle Prefix, P_CP_t(``i")P_l, 3 \rangle$ \\ $\langle Prefix, P_CP_t(``in")P_l, 3 \rangle$ \\ $\langle Prefix, P_t(``Din")P_l, 3 \rangle$} \\ \hline
"Dinos " $\rightarrow$ "Din" & \tabincell{c}{$\langle Prefix, P_t(``D")P_lP_s, 3 \rangle$ \\$\langle Prefix, P_CP_lP_s, 3 \rangle$ \\ $\langle Prefix, P_CP_t(``i")P_lP_s, 3 \rangle$ \\ $\langle Prefix, P_CP_t(``in")P_lP_s, 3 \rangle$ \\ $\langle Prefix, P_t(``Din")P_lP_s, 3 \rangle$}\\ \hline
"Dinos in" $\rightarrow$ "Din" & \tabincell{c}{$\langle Prefix, P_t(``D")P_lP_sP_l, 3 \rangle$ \\$\langle Prefix, P_CP_lP_sP_l, 3 \rangle$ \\ $\langle Prefix, P_CP_t(``i")P_lP_sP_l, 3 \rangle$ \\ $\langle Prefix, P_CP_t(``in")P_lP_sP_l, 3 \rangle$ \\ $\langle Prefix, P_t(``Din")P_lP_sP_l, 3 \rangle$}\\ \hline
"Dinos in " $\rightarrow$ "Din" & \tabincell{c}{$\langle Prefix, P_t(``D")P_lP_sP_lP_s, 3 \rangle$ \\$\langle Prefix, P_CP_lP_sP_lP_s, 3 \rangle$ \\ $\langle Prefix, P_CP_t(``i")P_lP_sP_lP_s, 3 \rangle$ \\ $\langle Prefix, P_CP_t(``in")P_lP_sP_lP_s, 3 \rangle$ \\ $\langle Prefix, P_t(``Din")P_lP_sP_lP_s, 3 \rangle$}\\ \hline
"Dinos in Kas" $\rightarrow$ "Din" & \tabincell{c}{$\langle Prefix, P_t(``D")P_lP_sP_lP_sP_CP_l, 3 \rangle$ \\$\langle Prefix, P_CP_lP_sP_lP_sP_CP_l, 3 \rangle$ \\ $\langle Prefix, P_CP_t(``i")P_lP_sP_lP_sP_CP_l, 3 \rangle$ \\ $\langle Prefix, P_CP_t(``in")P_lP_sP_lP_sP_CP_l, 3 \rangle$ \\ $\langle Prefix, P_t(``Din")P_lP_sP_lP_sP_CP_l, 3 \rangle$}\\ \hline
\end{tabular}
\vspace{-2em}
\center
\end{table}

\begin{table}[!t]\vspace{1em}
\begin{adjustwidth}{0cm}{}
\setlength{\tabcolsep}{1pt}
\small
\center
\caption{Candidate rules for "(2002-06-29)" $\rightarrow$ "\%06\%"\label{table:rules:29}}
\vspace{-1em}
\begin{tabular}{|c|c|c|c|}
\hline
& Rules  \\ \hline
"06" $\rightarrow$ "06" & \tabincell{c}{$\langle Prefix, P_t(``06"), 2 \rangle$ \\ $\langle Prefix, P_nP_t(``6"), 2 \rangle$ \\ $\langle Prefix, P_n, 2 \rangle$ \\ $\langle Prefix, P_t(``0")P_n, 2 \rangle$} \\ \hline
"06-" $\rightarrow$ "06" & \tabincell{c}{$\langle Prefix, P_t(``06")P_t(``\mathcal{-}"), 2 \rangle$ \\ $\langle Prefix, P_nP_t(``6")P_t(``\mathcal{-}"), 2 \rangle$ \\ $\langle Prefix, P_nP_t(``\mathcal{-}"), 2 \rangle$ \\ $\langle Prefix, P_t(``0")P_nP_t(``\mathcal{-}"), 2 \rangle$} \\ \hline
"06-29" $\rightarrow$ "06" & \tabincell{c}{$\langle Prefix, P_t(``06")P_t(``\mathcal{-}")P_n, 2 \rangle$ \\ $\langle Prefix, P_nP_t(``6")P_t(``\mathcal{-}")P_n, 2 \rangle$ \\ $\langle Prefix, P_nP_t(``\mathcal{-}")P_n, 2 \rangle$ \\ $\langle Prefix, P_t(``0")P_nP_t(``\mathcal{-}")P_n, 2 \rangle$}\\ \hline
"06-29)" $\rightarrow$ "06" & \tabincell{c}{$\langle Prefix, P_t(``06")P_t(``\mathcal{-}")P_nP_t(``)"), 2 \rangle$ \\ $\langle Prefix, P_nP_t(``6")P_t(``\mathcal{-}")P_nP_t(``)"), 2 \rangle$ \\ $\langle Prefix, P_nP_t(``\mathcal{-}")P_nP_t(``)"), 2 \rangle$ \\ $\langle Prefix, P_t(``0")P_nP_t(``\mathcal{-}")P_nP_t(``)"), 2 \rangle$}\\ \hline
"06" $\rightarrow$ "06" & \tabincell{c}{$\langle Suffix, P_t(``06"), 2 \rangle$ \\ $\langle Suffix, P_nP_t(``6"), 2 \rangle$ \\ $\langle Suffix, P_n, 2 \rangle$ \\ $\langle Suffix, P_t(``0")P_n, 2 \rangle$}\\ \hline
"-06" $\rightarrow$ "06" & \tabincell{c}{$\langle Suffix, P_t(``\mathcal{-}")P_t(``06"), 2 \rangle$ \\ $\langle Suffix, P_t(``\mathcal{-}")P_nP_t(``6"), 2 \rangle$ \\ $\langle Suffix, P_t(``\mathcal{-}")P_n, 2 \rangle$ \\ $\langle Suffix, P_t(``\mathcal{-}")P_t(``0")P_n, 2 \rangle$}\\ \hline
"2002-06" $\rightarrow$ "06" & \tabincell{c}{$\langle Suffix, P_nP_t(``\mathcal{-}")P_t(``06"), 2 \rangle$ \\ $\langle Suffix, P_nP_t(``\mathcal{-}")P_nP_t(``6"), 2 \rangle$ \\ $\langle Suffix, P_nP_t(``\mathcal{-}")P_n, 2 \rangle$ \\ $\langle Suffix, P_nP_t(``\mathcal{-}")P_t(``0")P_n, 2 \rangle$}\\ \hline
"(2002-06" $\rightarrow$ "06" & \tabincell{c}{$\langle Suffix, P_t(``(")P_nP_t(``\mathcal{-}")P_t(``06"), 2 \rangle$ \\ $\langle Suffix, P_t(``(")P_nP_t(``\mathcal{-}")P_nP_t(``6"), 2 \rangle$ \\ $\langle Suffix, P_t(``(")P_nP_t(``\mathcal{-}")P_n, 2 \rangle$ \\ $\langle Suffix, P_t(``(")P_nP_t(``\mathcal{-}")P_t(``0")P_n, 2 \rangle$}\\ \hline

\end{tabular}
\vspace{-2em}
\center
\end{adjustwidth}
\end{table}

\vspace{-.5em}

\subsection{Rule Generation}
\label{sec:string:extract}

\hi{Rule Definition. } Each rule can be expressed as a program. We borrow the idea from domain specific language (DSL) proposed by Gulwani~\cite{DBLP:conf/popl/Gulwani11,DBLP:journals/cacm/GulwaniHS12} to define a rule, which is composed of three parts, {\it pattern, string function and size}. The pattern matches substrings of tuples in the dataset. The string function decides which substring of tuples should be extracted. The size indicates the length of the substring to be extracted.  

The pattern includes capital letters $\mathcal{P}_C$, lowercase letters $\mathcal{P}_l$, numerical values $\mathcal{P}_n$, white spaces $\mathcal{P}_s$ and exact matching token $\mathcal{P}_t(T)$ which can only match a specific substring $T$. The string function includes two types, {\it Prefix} and {\it Suffix}. Prefix extracts the prefix of the string and Suffix extracts the suffix of the string. The rule can be formalized as below:
\begin{equation*}
\begin{split}
&\mathcal{F} \in \{\text{Prefix, Suffix}\} \\
&\mathcal{P} \in combination\{\mathcal{P}_C, \mathcal{P}_l,\mathcal{P}_s,\mathcal{P}_n,\mathcal{P}_t(T)\} \\
&\qquad\mathcal{P}_C = [A\mathcal{-}Z]+ \\
&\qquad\mathcal{P}_l = [a\mathcal{-}z]+ \\
&\qquad\mathcal{P}_s = whitespace+ \\
&\qquad\mathcal{P}_n = [0\mathcal{-}9]+ \\
&\qquad\mathcal{P}_t(T) = T\\
&rule = \langle \mathcal{F}, \mathcal{P}, \mathcal{L} \rangle
\end{split}
\end{equation*}
where $\mathcal{P}$ is a pattern, $\mathcal{F}$ is a string function, and $\mathcal{L}$ is the length of a substring.  

\hi{Rule Candidate Set. } Given a query string in the predicate and a tuple string in the dataset, we first find all substrings of the tuple string that match the query string. Then for each matched substring, we generate all possible patterns that map the query string to the substring, by enumerating all possible combinations of patterns in $\mathcal{P}_C$,  $\mathcal{P}_l$,  $\mathcal{P}_n$, $\mathcal{P}_s$, $\mathcal{P}_t(T)$. If the predicate is {\it prefix search} (e.g., LIKE ``Din\%''), then for each possible pattern $p$, we generate a rule (prefix, $p$, size of the query string).   If the predicate is {\it suffix search}, we generate a rule (suffix, $p$, size of the query string).  If the predicate is {\it substring search}, we generate a rule (prefix/suffix, $p$, size of the query string), based on how the query string matches the substring. All the possible rules will form a {\it Rule Candidate Set}.


For example, Table~\ref{table:rules:dk} shows some predicates and their sample substrings where `Dinos in Kas' is a value selected by the predicate "title LIKE `Din\%'".  Table~\ref{table:rules:29} shows the rules for ``\%06\%''.




\hi {Rule Selection.} Based on the candidate rules, we aim to find an {\it optimal} set of rules, which finds the minimum number of rules to cover the query workload. However, if we select those too general rules, the number of substrings would be too large. Therefore, we set an upper bound for the total number of extracted substrings.

Let $\mathcal{R}$  denote a subset of candidate rule set $C_\mathcal{R}$ which could cover the strings in the workload, $S_\mathcal{R}$ be the set of substrings which are extracted by rules in $\mathcal{R}$ from the datasets, and $S_\mathcal{W}$ is the set of strings in the workload. We aim to minimize the size of $\mathcal{R}$ with an upper bound $\mathcal{B}$ where $S_\mathcal{R}$ contains all strings in $S_\mathcal{W}$. The problem is formalized below:
\begin{equation*}
\begin{split}
&\mathcal{R} = arg\min_{\mathcal{R}\subseteq C_\mathcal{R}}(|\mathcal{R}|)\\
&\mathrm{ s.t. }\quad |S_\mathcal{R}| < \mathcal{B}, S_\mathcal{W} \subseteq S_\mathcal{R} \\
\end{split}
\end{equation*}

This is an NP-hard problem by a reduction from a classical set cover problem (SCP)\cite{scp}. Now the universe is $S_\mathcal{W}$, and the subset is $S_r \cap S_\mathcal{W}$ where $r \in C_\mathcal{R}$. We also have that the union of subsets $\sum_{r \in C_\mathcal{R}}S_r \cap S_\mathcal{W}$ equals to the universe $S_\mathcal{W}$. Our target is to find the minimum number of subsets to cover the universe.


We propose a greedy solution to address this problem approximately. We add a rule $r$ to the rule set $\mathcal{R}$ covering the most substrings in $S_\mathcal{W}$ each time. If the total size of $S_\mathcal{R}$ exceeds the bound $\mathcal{B}$, we remove the rule $r$ with the largest $S_r$ and repeat. The pseudo code of the algorithm is shown in Algorithm~\ref{algorithm:rule-select}.

For example, consider three rules in the Table~\ref{table:rules:29},  $\langle Prefix, P_t(``06"), 2 \rangle$, $\langle Prefix, P_n, 2 \rangle$ and $\langle Suffix, P_t(``(")P_nP_t(``\mathcal{-}")P_n, 2 \rangle$. The first rule can only extract `06'. The second rule can extract 6 substrings including \{`2002', `06', `29', `2014', `08', `26'\}. The third rule can extract \{`06', `08'\}. The third rule would be selected in our algorithm, because it's general and will not extract too many substrings.

\begin{algorithm}[!t]
\linesnumbered \SetVline
\caption{RuleSelection\label{algorithm:rule-select}}
\KwIn{Candidate rules set $C_\mathcal{R}$; An upper bound $B$; All string sets $S_r$ extracted by rules in $C_\mathcal{R}$ from datasets; Set of strings in the workload $S_\mathcal{W}$; }
\KwOut{Selected rule set $\mathcal{R}$}
Initialize rule set $\mathcal{R} = \emptyset$\;
\For{ r $\in$ $C_\mathcal{R}$ }{
$r$ $\rightarrow$ $\langle r, S_r, |S_r\mathcal{-}S_\mathcal{W}| \rangle$\;
}
Sort $C_\mathcal{R}$ in descending order of $|S_r\mathcal{-}S_\mathcal{W}|$\;
\While{ $S_\mathcal{W} - S_\mathcal{R} \neq \emptyset$ or $C_\mathcal{R} \neq \emptyset$ } {
$r^*$,$S_{r^*}$,$\mathcal{L}^*$ = $pop(C_\mathcal{R})$\;
$\mathcal{R}$ = add($\mathcal{R}$, $r^*$)\;
$S_\mathcal{R}$ = add($S_{r^*}$,$S_\mathcal{R}$)\;
\If{$|S_\mathcal{R}| \ge \mathcal{B}$} {
$r'$ = $arg\min_{r \in \mathcal{R}}\frac{|S_\mathcal{R} \cap S_\mathcal{W}|}{|S_\mathcal{R}|}$\;
$\mathcal{R}$ = remove($r'$,$\mathcal{R}$)\;
$S_\mathcal{R}$ = remove($S_{r'}$,$S_\mathcal{R}$)\;
}
}
return $\mathcal{R}$\;
\end{algorithm}

Based on the selected rules, we generate  all the string values in the dataset and store the extracted substrings in the dictionary.

\subsection{String Indexing}
\label{sec:string:structure}

\begin{figure}[!t]\vspace{1em}
\begin{center}
\centering
\vspace{-2.5em}
\includegraphics[width=0.45\textwidth]{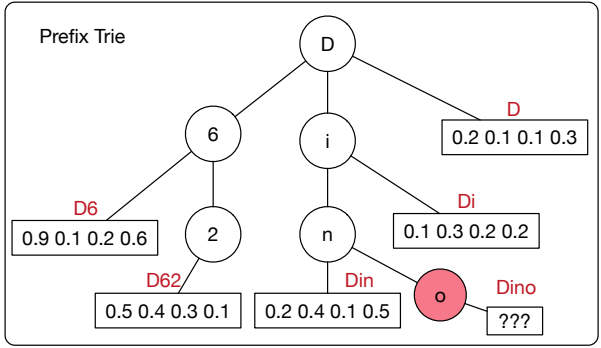}
\vspace{-1em}
\caption{Trie Index Structure\label{fig:trie-index}}
\end{center}
\vspace{-3em}
\end{figure}

There could be a large number of strings and it is expensive to maintain all strings in a dictionary.  In order to avoid storing a huge number of duplicate tokens, we build  a  trie index to store the mapping from a string to its code.  

\hi{String Indexing. } We use both prefix trie and suffix trie as string index. Figure~\ref{fig:trie-index} shows an example of a prefix trie.  Substrings extracted by the {\it prefix} function are stored into prefix trie and substrings extracted by the {\it suffix} function are stored in suffix trie. In this way, each string in the dictionary must have one or two paths in the index. Leaf nodes of the trie index are representation vectors of strings.

\hi{Online Searching.} When a new query comes, there may be some query strings which do not exist in the dictionary. These query strings may be in prefix searching (LIKE s\%), suffix searching(LIKE \%s), keyword searching(=) or containment searching(LIKE \%s\%). For prefix search, we search the longest prefix of the query  string. For suffix search, we search the longest suffix of the query string. For other searches, we search both the longest prefix and longest suffix of the query string, and then pick the longest one as the representation. Considering "title LIKE `Dino\%'", we search the prefix trie and take the representation of `Din' as the representation of `Dino'. In this way, we can encode queries online quickly.

%% file: src/exp.tex
\section{Experiments} \label{sec:exp}

We conduct extensive experiments to evaluate our method from three aspects. (1) The effectiveness of our tree-structured model on cost and cardinality estimation. (2) The effectiveness of predicates embedding. (3) The efficiency of our model on cost estimation. 


\subsection{Experiment Setting}

\hi{Datasets.} We use the real dataset IMDB and the real workload JOB. It is much harder to estimate the cardinality and cost on the IMDB dataset than TPC-H,  because of the correlations and skew distributions of the real-world data. The IMDB dataset includes 22 tables, which are joined on primary keys and foreign keys. We build indexes on primary keys. We use two types of query workloads. 

 (1) The first workload contains  predicates with numeric attributes only~\cite{DBLP:conf/cidr/KipfKRLBK19}. It contains synthesis workload, scale workload and JOB-light workload\footnote{https://github.com/andreaskipf/learnedcardinalities}. We adopt all three workloads with only numeric predicates. The {\it Synthesis} workload contains queries with 2 joins at most and there are 5000 queries. The {\it Scale} workload contains queries with 0-4 joins and there are 500 queries. The {\it JOB-light} workload contains queries with 1-4 joins, and there are 70 queries.

(2) The second workload contains complex predicates with string attributes. We generate training data based on the join graph of IMDB and predicates used in the JOB workloads. We take  90\% of generated queries as training data and 10\% as validation data. The 113 JOB queries\footnote{https://github.com/gregrahn/join-order-benchmark} are taken as the test workload. Different from previous cardinality estimation methods \cite{DBLP:conf/cidr/KipfKRLBK19}, we obtain query plans from PostgreSQL, and use the plans to train   our model.

\hi{Metrics.}
The $mean$ is the average errors of all the tested queries. The $max$ is the maximum errors in the tested workload. The $median$ is the median of errors of all the tested queries. The $\mathcal{K}th$  is the average of top-\{1-$\mathcal{K}\%$\} largest errors in the tested workload.

\hi{Environment.} We use a machine with Intel(R) Xeon(R) CPU E5-2630 v4, 128GB Memory, and GeForce GTX 1080.

\begin{table}[!t]
\setlength{\tabcolsep}{3pt}
\small
\begin{adjustwidth}{0cm}{}
\caption{Methods on numeric workloads \label{table:method:numeric}}
\vspace{-1em}
\begin{tabular}{|c|c|c|c|c|c|}
\hline
Methods & Target & Represent & Predicate & Estimate & Sample\\ \hline
\pgcard & Card & No & No & No & No \\ \hline
\mscncard & Card & No & No & SING & Yes\\ \hline
\mscnnosamplecard & Card & No & No &SING & No\\ \hline
\tlstmnosamplecard & Card & LSTM & LSTM &SING & No\\ \hline
\tnncard & Card & NN & LSTM & SING & Yes\\ \hline
\tlstmcard & Card & LSTM & LSTM & SING & Yes\\ \hline
\pgcost & Cost & No & No & No & No\\ \hline
\mscncost & Cost & No & No & SING & Yes\\ \hline
\tlstmcostonly & Cost & LSTM & LSTM &SING & Yes\\ \hline
\tnncost & Cost & NN & LSTM &MULT & Yes\\ \hline
\tlstmcost & Cost & LSTM & LSTM &MULT& Yes\\ \hline
\end{tabular}
\end{adjustwidth}
\vspace{-.5em}

\setlength{\tabcolsep}{5pt}
\small
\begin{adjustwidth}{0em}{}
\caption{Cardinality errors on numeric workloads \label{table:card:numeric}}
\vspace{-0em}
\begin{tabular}{|c|c|c|c|c|c|c|}
\hline
\textbf{JOB-light} & median & 90th & 95th & 99th & max & mean \\ \hline
\pgcard & 7.93 & 164 & 1104 & 2912 & 3477 & 174  \\\hline
\mscncard & 3.82 & 78.4 & 362 & 927 & 1110 & 57.9 \\\hline
\tnncard & \textbf{2.95} & 76.8 & 275 & 799 & 902 & 49.8 \\\hline
\tlstmcard & 3.73 & \textbf{50.8} & \textbf{157} & \textbf{256} & \textbf{289} & \textbf{24.9} \\\hline
\hline
\textbf{Synthetic} & median & 90th & 95th & 99th & max & mean \\ \hline
\pgcard & 1.69 & 9.57 & 23.9 & 465 & 373901 & 154 \\\hline
\mscncard & \textbf{1.18} & 3.32 & 6.84 & 30.51 & 1322 & 2.89\\\hline
\tnncard & 1.40 & 5.51 & 10.7 & 43.1 & 441 & 3.57  \\\hline
\tlstmcard & 1.20 & \textbf{3.21} & \textbf{6.12} & \textbf{25.2} & \textbf{357} & \textbf{2.87}  \\\hline
\hline
\textbf{Scale} & median & 90th & 95th & 99th & max & mean\\ \hline
\pgcard & 2.59 & 200 & 540 & 1816 & 233863 & 568  \\\hline
\mscncard & \textbf{1.42} & \textbf{37.4} & 140 & 793 & 3666 & 35.1\\\hline
\tnncard & 1.59 & 58.7 & 141 & 573 & 2238 & 31.3   \\\hline
\tlstmcard & 1.43 & 38.8 & \textbf{139} & \textbf{469} & \textbf{1892} & \textbf{28.1} \\\hline
\end{tabular}
\end{adjustwidth}
\vspace{-1em}

\small
\begin{adjustwidth}{0em}{}
\caption{Cost errors on numeric workloads \label{table:cost:numeric}}
\vspace{-0em}
\begin{tabular}{|c|c|c|c|c|c|c|}
\hline
\textbf{JOB-light} & median & 90th & 95th & 99th & max & mean\\ \hline
\pgcost & 26.8 & 332 & 696 & 2740 & 3020 & 173\\\hline
\mscncost & 4.75 & 11.3 & 40.1 & 563 & 987 & 27.4\\\hline
\tlstmcostonly & 3.66 & 32.1 & 80.3 & 445 & 583 & 17\\\hline
\tnncost & 2.06 & 25.5 & 134 & 293 & 401 & 19.1\\\hline
\tlstmcost & \textbf{1.85} & \textbf{13.2} & \textbf{22.9} & \textbf{95} & \textbf{123} & \textbf{5.81}\\\hline
\hline
\textbf{Synthetic} & median & 90th & 95th & 99th & max & mean \\ \hline
\pgcost & 15.1 & 65.1 & 173 & 1200 & 8040 & 62.7\\\hline
\mscncost & 3.14 & 7.43 & 18.1 & 65.8 & 739 & 10.3\\\hline
\tlstmcostonly & 1.56 & 4.47 & 10.7 & 57.7 & 689 & 4.45\\\hline
\tnncost & \textbf{1.49} & 4.50 & 10.6 & 61.5 & 718 & 4.35\\\hline
\tlstmcost & \textbf{1.49} & \textbf{4.33} & \textbf{10.2} & \textbf{55.8} & \textbf{624} & \textbf{4.16}\\\hline
\hline
\textbf{Scale} & median & 90th & 95th & 99th & max & mean\\ \hline
\pgcost & 13.3 & 38.9 & 81.1 & 718 & 1473 & 35.7 \\\hline
\mscncost & 1.79 & 10.6 & 27.1 & 88.8 & 1027 & 8.22 \\\hline
\tlstmcostonly & 1.58 & 5.51 & 14.4 & 70.1 & 611 & 5.21\\\hline
\tnncost & 1.61 & 5.37 & 13.5 & 72.7 & 714 & 5.53\\\hline
\tlstmcost & \textbf{1.56} & \textbf{5.56} & \textbf{12.2} & \textbf{68.6} & \textbf{254} & \textbf{4.41} \\\hline

\end{tabular}
\end{adjustwidth}
\vspace{-2em}
\end{table}

\subsection{Effectiveness on workloads with only numeric predicates}

\hi{Methods.} Table~\ref{table:method:numeric} shows the methods evaluated on workloads with numeric predicates. The {\it Target} is cardinality or cost. The {\it Represent} is the model we use as representation layer described in Section~\ref{sec:model:represent}, including LSTM and Neural Network. The {\it Predicate} is the model we use as a predicate embedding layer described in Section~\ref{sec:model:embedding}, including Min-Max Pooling model and tree-LSTM. The {\it Estimate} is the model we use as an estimation layer described in Section~\ref{sec:model:estimate}, and it indicates whether using multitask training or using cost estimation only. The {\it Sample} indicates whether the sample bitmap is used. 

\begin{figure}[!t]
\begin{subfigure}{.235\textwidth}
  \centering
  \includegraphics[width=1\linewidth]{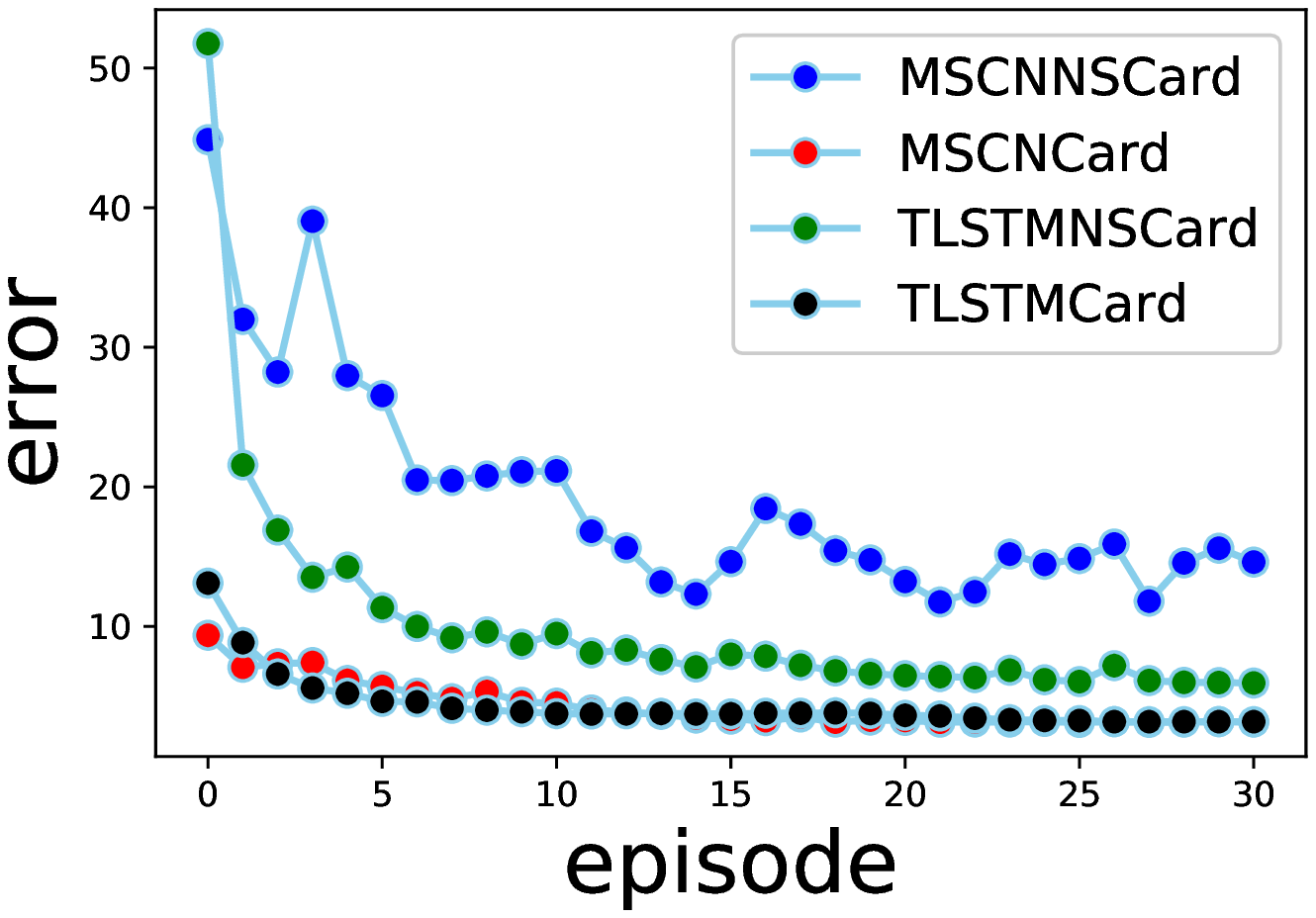}
\caption{Card Validation Error}\label{fig:validate:card}
\end{subfigure}
\begin{subfigure}{.235\textwidth}
  \centering
  \includegraphics[width=1\linewidth]{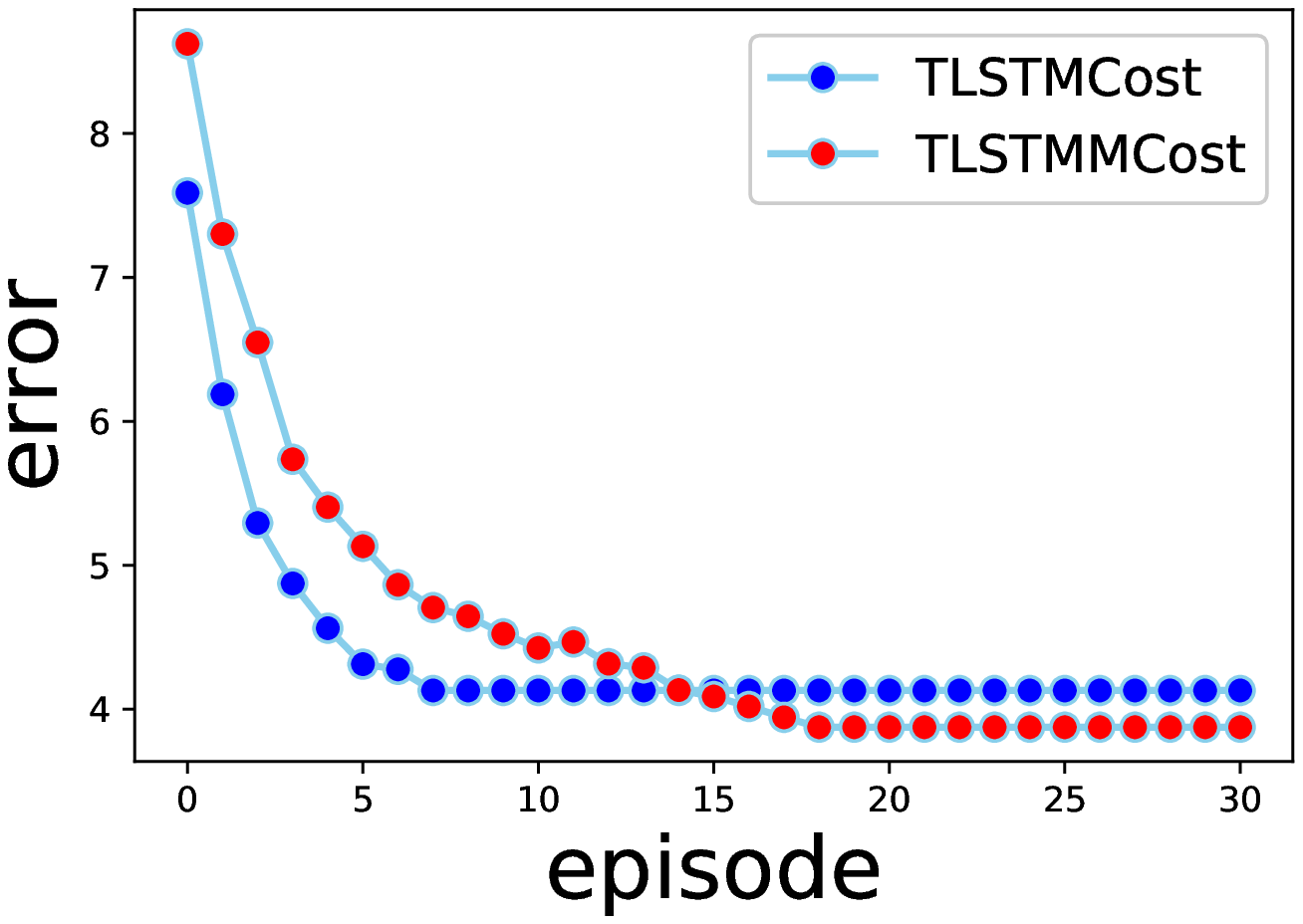}
\caption{Cost Validation Error}\label{fig:validate:cost}
\end{subfigure}
\vspace{-1em}
\caption{Validation Error on Numeric Workload}\vspace{-2em}
\end{figure}

As existing methods only support the predicates with numerical values only, we report the results of different methods running on the workload which is composed of queries with numeric predicates only. The validation error is shown in Figure~\ref{fig:validate:card} and Figure~\ref{fig:validate:cost}. After models are converged, we test them on the workload including \textit{Synthetic}, \textit{Scale} and \textit{JOB-light}, and report the results in Tables~\ref{table:card:numeric} and \ref{table:cost:numeric}.


\hi{Sample vs Non-Sample.} As shown in Figure~\ref{fig:validate:card}, the methods with sample bitmap (e.g., \tlstmcard and \mscncard) outperform the methods without bitmap (e.g., \tlstmnosamplecard and \mscnnosamplecard) on the validation data, because sample bitmap reveals both the distribution of data and the semantic of predicates explicitly, and the model could easily extract the approximate intermediate result distribution from the bitmap.

\hi{Tree-structured Model vs CNN.}  \tlstmcard outperforms \mscncard, because the tree-structured model could capture the query predicate and learn the semantic of them much better than convolutional network. However, \mscncard and \tlstmcard make no much difference on validation queries because the validation workload contains only easy queries like the training workload, and the sample bitmap of length 1000 is sufficient to represent distribution of each intermediate result and brings little error for the cardinality estimation of queries. Thus the advantage of tree-structured model is not significant on easy queries for cardinality estimation. However, if we apply the model to harder queries like the JOB-light workload, then small bias or 0-tuple problem of the sample bitmap would lead to large errors of the estimated cardinality. On the JOB-light workload, \tnncard with tree-structured model outperforms \mscncard by 20\% on mean error for cardinality estimation, and \tnncost outperforms \mscncost by 40\% on mean error for cost estimation. On the Scale workload, \tnncard outperforms \mscncard by 2 times on max error for cardinality estimation and \tnncost outperforms \mscncost by 2 times on mean error for cost estimation. On Synthetic workload, \tnncard outperforms \mscncard by 4 times on max error for cardinality estimation and \tnncost outperforms \mscncost by 3 times on max error. The main reasons are three-fold. (1) Tree-structured model is good at representing complicated plans and predicates. (2) Tree-structured model captures more correlations for complex queries. (3) Cost model used by \pgcost and \mscncost is hard to be tuned.

\hi{LSTM vs NN.} On the JOB-light workload, \tlstmcard outperforms \tnncard by 3 times on max error for cardinality estimation and 2 times on the mean error. \tlstmcost outperforms \tnncost by 3 times on mean error and max error for cost estimation. On the Synthetic workload, \tlstmcard outperforms \tnncard on all the cardinality errors. On the Scale workload, tree-LSTM outperforms tree-NN on all the cardinality errors and cost errors. \tlstmcard outperforms \tnncard on harder queries for cardinality estimation, and \tlstmcost outperforms \tnncost on all the cost errors. These results show that LSTM can help the representation model to learn more robust representations for sub-plans, because the LSTM has an extra channel to avoid information vanishing for complex queries.

\hi{Multi-Learning vs Cost Only Learning.} On the JOB-light workload, multitask technique can help the model to achieve 3 times improvement on 90-99th, mean and max error for cost estimation. On the Synthetic workload, \tlstmcost outperforms \tlstmcostonly on all the cost errors. On the Scale workload, \tlstmcost outperforms \tlstmcostonly by more than 2 times on max error for cost estimation. The reason is that multitask learning can improve the generalization ability of the model for complex queries.

In summary, sample bitmap, tree-structured model, LSTM, and multitask learning can improve the quality of cost and cardinality estimation.

\subsection{Effectiveness on the JOB workload with both string and numeric predicates}

\hi{Methods.} Table~\ref{table:method:job} shows the methods tested on the JOB workload with string and numeric predicates.  
The {\it String} shows methods in Section~\ref{sec:embedding} which use the string embedding technique, including embedding on rules generation strings and embedding only on string values in datasets.

\begin{table}[!t]
\setlength{\tabcolsep}{.5pt}
\small
\vspace{0em}
\caption{Methods on JOB workload (with strings) \label{table:method:job}}
\vspace{-1em}
\begin{adjustwidth}{0em}{}
\begin{tabular}{|c|c|c|c|c|c|c|}
\hline
Methods & Target & Represent & Predicate & Estimate & String \\ \hline
\pgcard & Card & No & No & No &  No\\ \hline
\tlstmhashcard & Card & LSTM & LSTM & MULT & Hash\\ \hline
\tlstmembwcard & Card & LSTM & LSTM & MULT & Embed\\ \hline
\tlstmembtcard & Card & LSTM & LSTM & MULT & Rule+Embed\\ \hline
\tpembtcard & Card & LSTM & Pool & MULT & Rule+Embed\\ \hline
\pgcost & Cost & No & No & No & No \\ \hline
\tlstmhashcost & Cost & LSTM & LSTM & MULT & Hash\\ \hline
\tlstmembwcost & Cost & LSTM & LSTM & MULT & Embed\\ \hline
\tlstmembtcost & Cost & LSTM & LSTM & MULT & Rule+Embed\\ \hline
\tpembtcost & Cost & LSTM & Pool & MULT & Rule+Embed\\ \hline
\end{tabular}
\end{adjustwidth}
\vspace{-.5em}
\setlength{\tabcolsep}{3pt}
\small
\begin{adjustwidth}{0em}{}
\caption{Cardinality errors on the JOB workload \label{table:card:job}}
\vspace{-1em}
\begin{tabular}{|c|c|c|c|c|c|c|}
\hline
\textbf{Cardinality} & median & 90th & 95th & 99th & max & mean \\ \hline
\pgcard & 184 & 8303 & 34204 & 1.06e5 & 6.70e5 & 10416 \\\hline
\tlstmhashcard & 11.1 & 207 & 359 & 824 & 1371 & 83.3  \\\hline
\tlstmembwcard & 11.6 & 181 & 339 & 777 & 1142 & 70.2 \\\hline
\tlstmembtcard & 10.9 & 136 & 227 & 682 & 904 & 55.0\\\hline
\tpembtcard & \textbf{10.1} & \textbf{74.7} & \textbf{193} & \textbf{679} & \textbf{798} & \textbf{47.5} \\\hline
\end{tabular}
\end{adjustwidth}
\vspace{-.5em}
\setlength{\tabcolsep}{3pt}
\small
\begin{adjustwidth}{0em}{}
\caption{Cost errors on the JOB workload \label{table:cost:job}}
\vspace{-1em}
\begin{tabular}{|c|c|c|c|c|c|c|}
\hline
\textbf{Cost} & median & 90th & 95th & 99th & max & mean \\ \hline
\pgcost & 4.90 & 80.8 & 104 & 3577 & 4920 & 105 \\\hline
\tlstmhashcost & 4.47 & 53.6 & 149 & 239 & 478 & 24.1  \\\hline
\tlstmembwcost & 4.12 & 18.1 & 44.1 & 105 & 166 & 10.3\\\hline
\tlstmembtcost & 4.28 & 13.3 & 22.5 & 104 & 126 & 8.6 \\\hline
\tpembtcost & \textbf{4.07} & \textbf{11.6} & \textbf{17.5} & \textbf{63.1} & \textbf{67.3} & \textbf{7.06} \\\hline
\end{tabular}
\end{adjustwidth}
\vspace{-2em}
\end{table}

In order to investigate different techniques proposed in the paper effectively, we divide the training data into two parts and train on them respectively. The first is workload without join, and the second  is workload with multiple joins. All these training queries contain complicated predicates on both numeric and string values, and they are generated randomly. Firstly, we train our models on the first workload and test them on single table validation workload, and this can compare the performance of different predicate embedding techniques directly. Secondly, we train our models on the second workload and evaluate them on 113 JOB queries,  and this can compare the effects of different predicate embedding techniques on complicated queries estimation. 


\subsubsection{Evaluation on single table workload} 

\begin{figure}[!t]\vspace{1em}
\begin{center}
\centering
\vspace{-2.5em}
\includegraphics[width=0.4\textwidth]{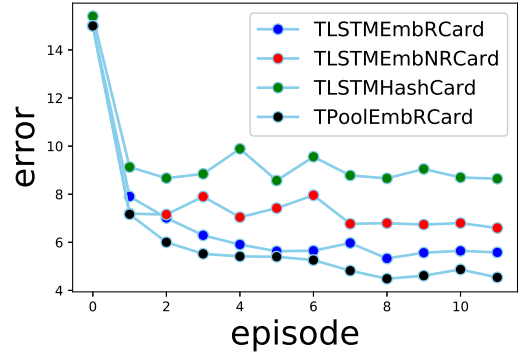}
\vspace{-1em}
\caption{Cardinality Validation Error on A Single Table\label{fig:validate:single:card}}
\end{center}
\vspace{-2em}
\end{figure}

The predicates in the workload contain string equal search, string pattern search, range query and numeric equal search. For conjunction predicates, the complex predicates are composed of expressions with `AND' and `OR' semantics. The most complex  predicate in the workload has 4 boolean conjunctions and 5 expressions. We set the batch size as 64 and divide all 56,000 queries into 882 batches. We take the first 800 batches as the training data, and the remainders as validation data. We use the Adam optimizer and the learning rate is 0.001. Since the semantic of predicates doesn't have much effect on execution cost on single table queries (Scan operation on the same table takes similar time no matter what predicate is.), we only report the validation error for cardinality estimation. The results are shown in Figure~\ref{fig:validate:single:card}.

\hi {Hash Bitmap vs String Embedding. } As shown in Figure~\ref{fig:validate:single:card}, \tlstmhashcard performs the worst on validation queries and it converge speed is the slowest. \tlstmembwcard outperforms \tlstmhashcard by 30\%, because string embedding can capture coexistence relation among different strings to improve the performance on the validation workload.

\hi {Rule vs No-Rule. } In Figure~\ref{fig:validate:single:card},  \tlstmembtcard outperforms \tlstmembwcard by around 15\% on cardinality estimation, because the coverage of strings in the workload to be trained is different, and the rules are selected for covering more strings in the workload and the method with rule can pre-train many more strings and encode them with more accurate distributed representations.

\hi {Tree-Pooling Predicate vs Tree-LSTM Predicate. } Both \tlstmembtcard and \tpembtcard can make full use of string embedding, but they make difference in cardinality estimation (\tpembtcard outperforms \tlstmembtcard by 20\% on validation workload), because the tree structure with Min-Max pooling is more capable of representing compound predicate in semantic, and it can learn a better predicate representation with stronger generalization ability.

\begin{figure}[!t]
    \centering
    \begin{subfigure}[h]{0.235\textwidth}
        \includegraphics[width=\textwidth]{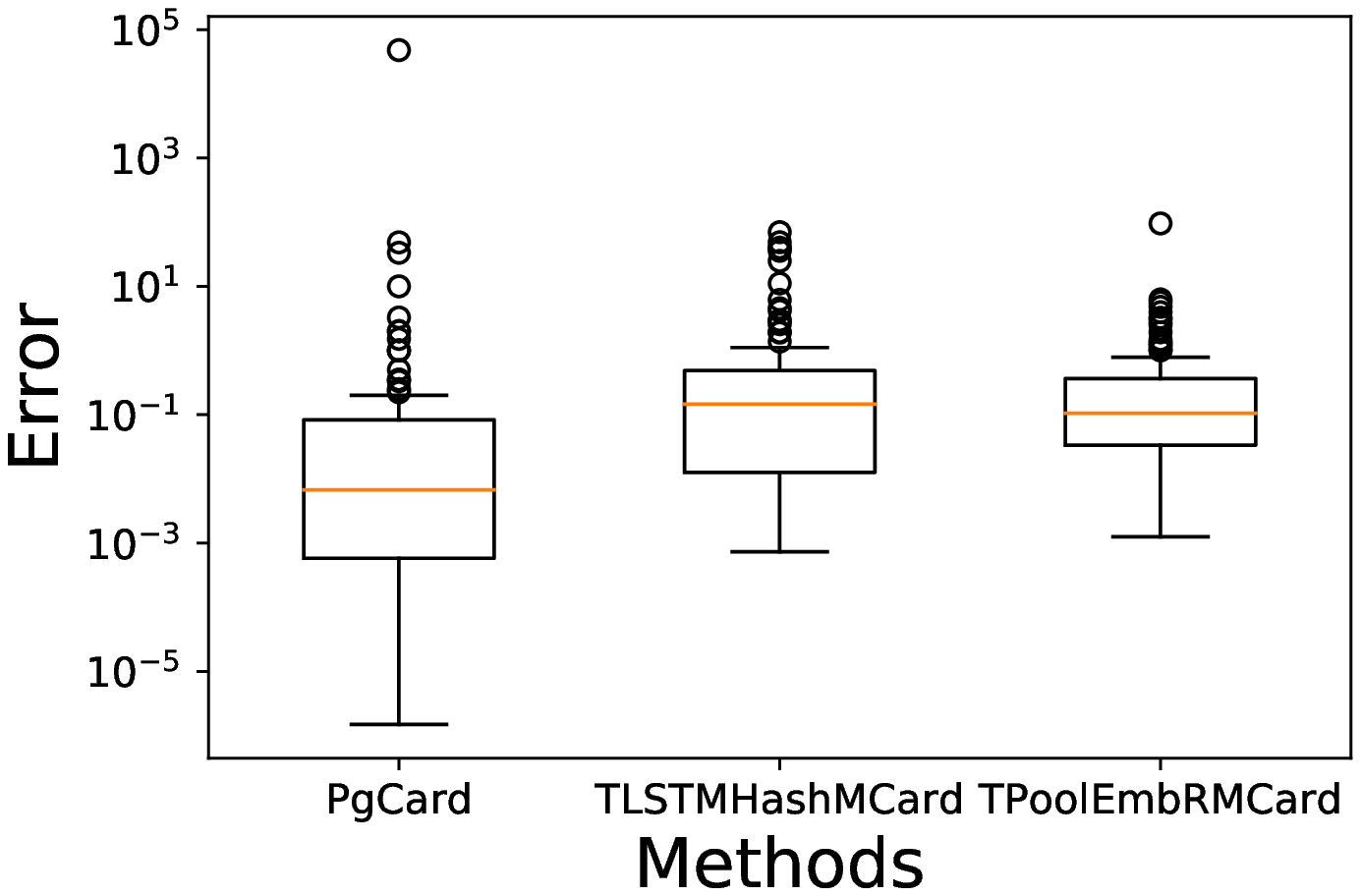}
        \caption{Cardinality}
        \label{subfig:range:card}
    \end{subfigure}
    \begin{subfigure}[h]{0.235\textwidth}
        \includegraphics[width=\textwidth]{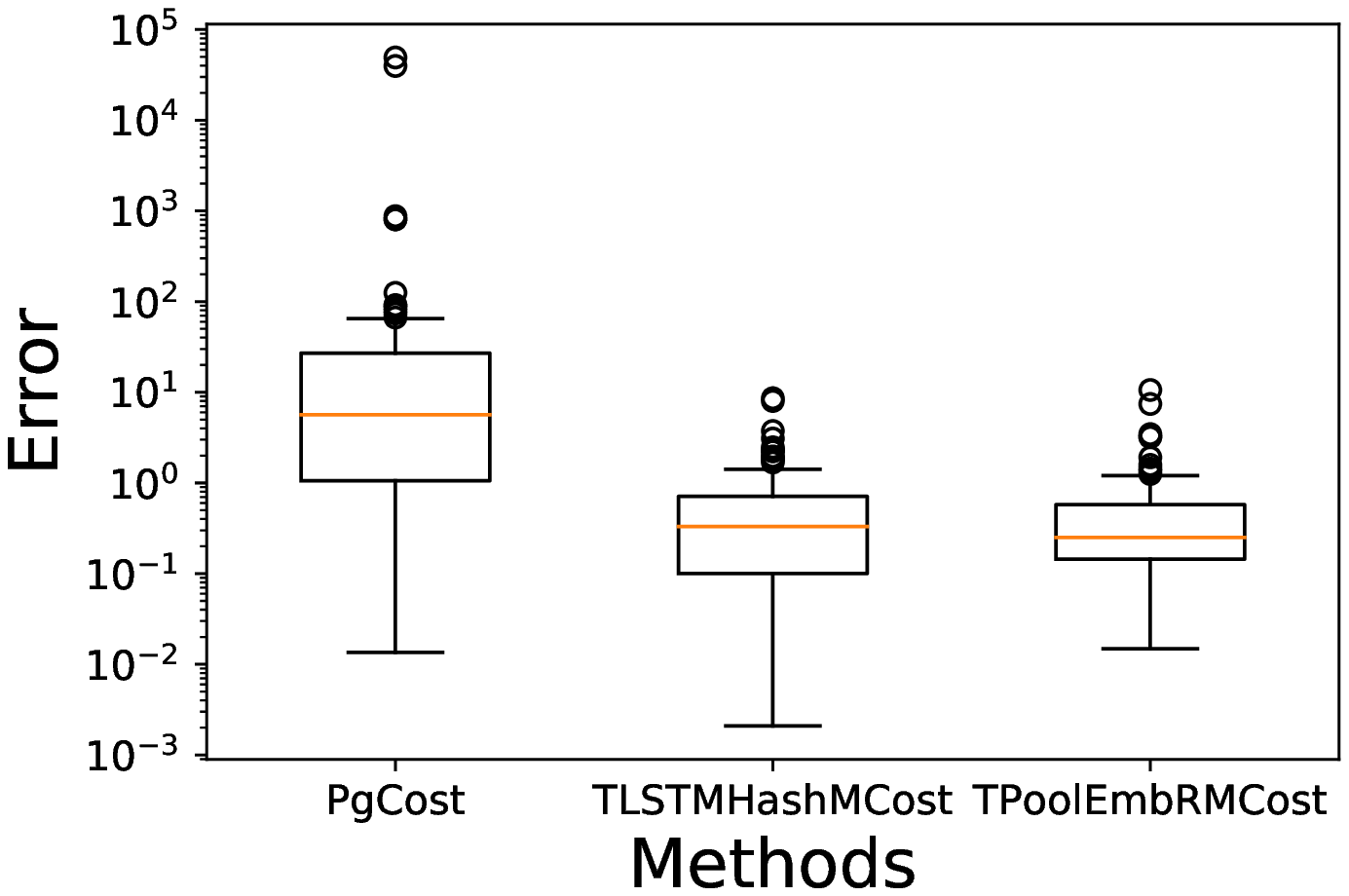}
        \caption{Cost}
        \label{subfig:range:card}
    \end{subfigure}
    \vspace{-1em}
    \caption[]{Estimation errors on the JOB workload. The box boundaries are at the 25th/50th/75th percentiles\label{fig:job:range}}
    \vspace{-2em}
\end{figure}

\begin{figure*}[!t]
    \centering
    \begin{subfigure}[h]{0.31\textwidth}
        \includegraphics[width=\textwidth]{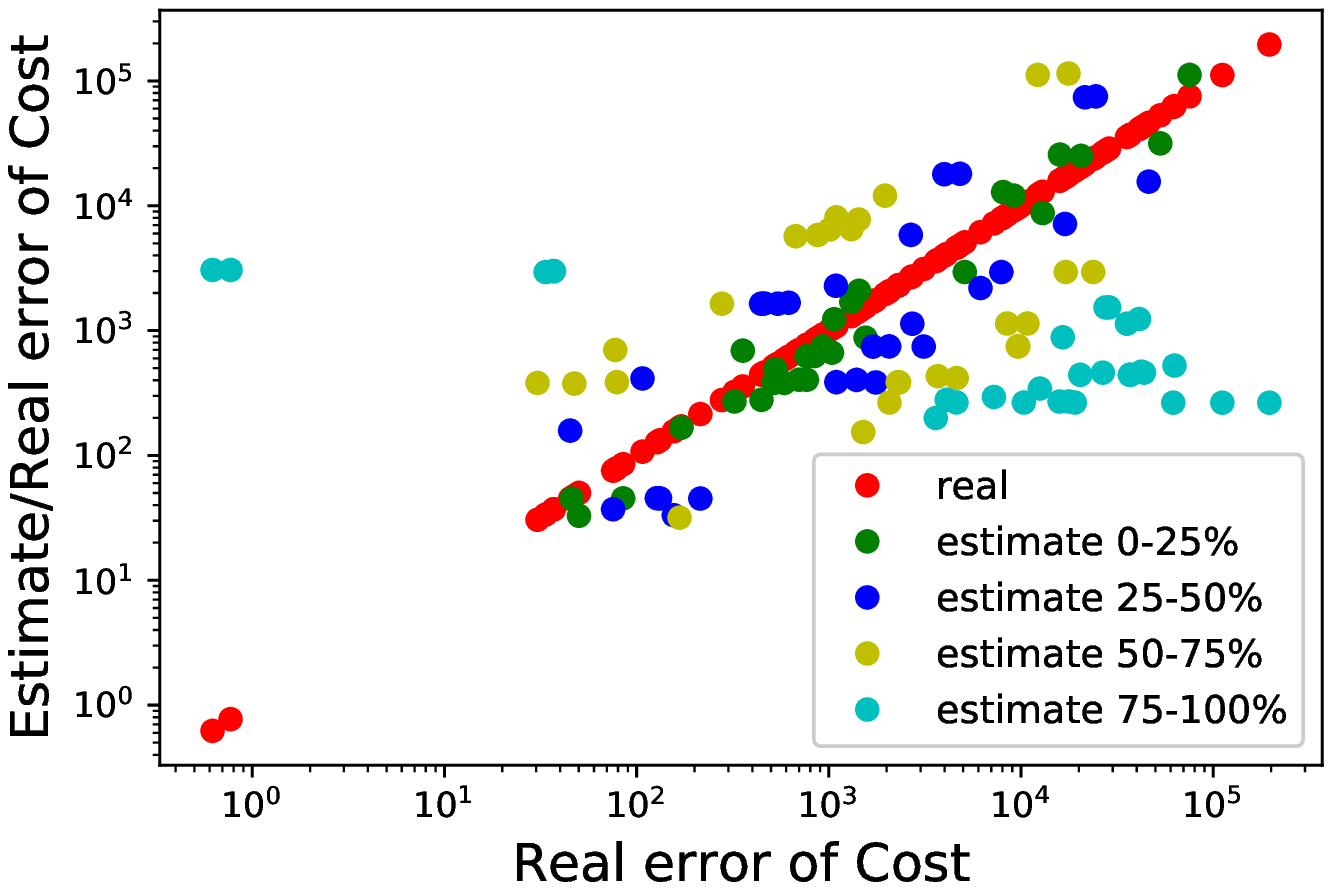}
        \caption{\pgcost}
        \label{subfig:search_review_sizes}
    \end{subfigure}
    \begin{subfigure}[h]{0.31\textwidth}
        \includegraphics[width=\textwidth]{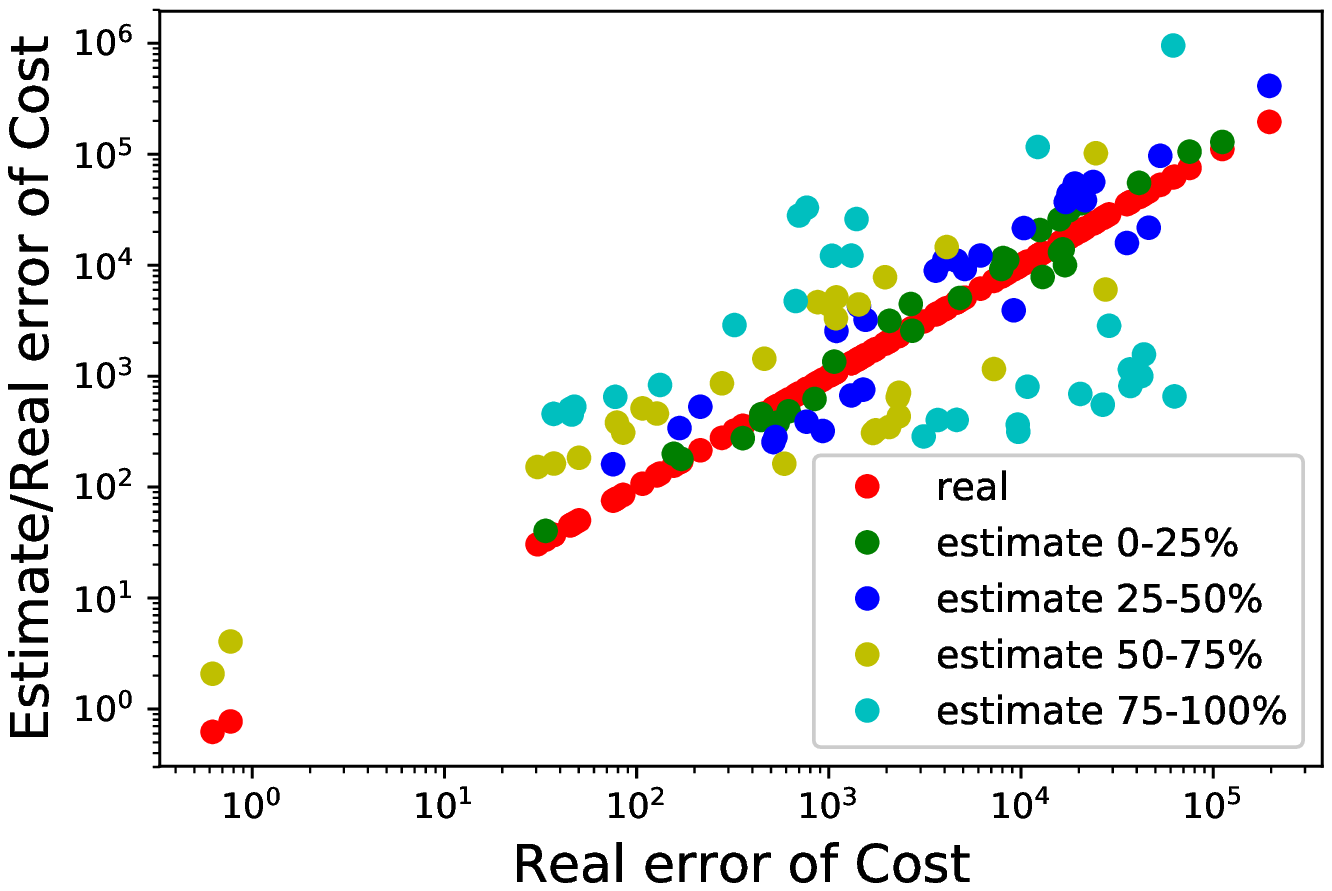}
        \caption{\tlstmembwcost}
        \label{subfig:search_review_cores}
    \end{subfigure}
    \begin{subfigure}{0.31\textwidth}
        \includegraphics[width=\textwidth]{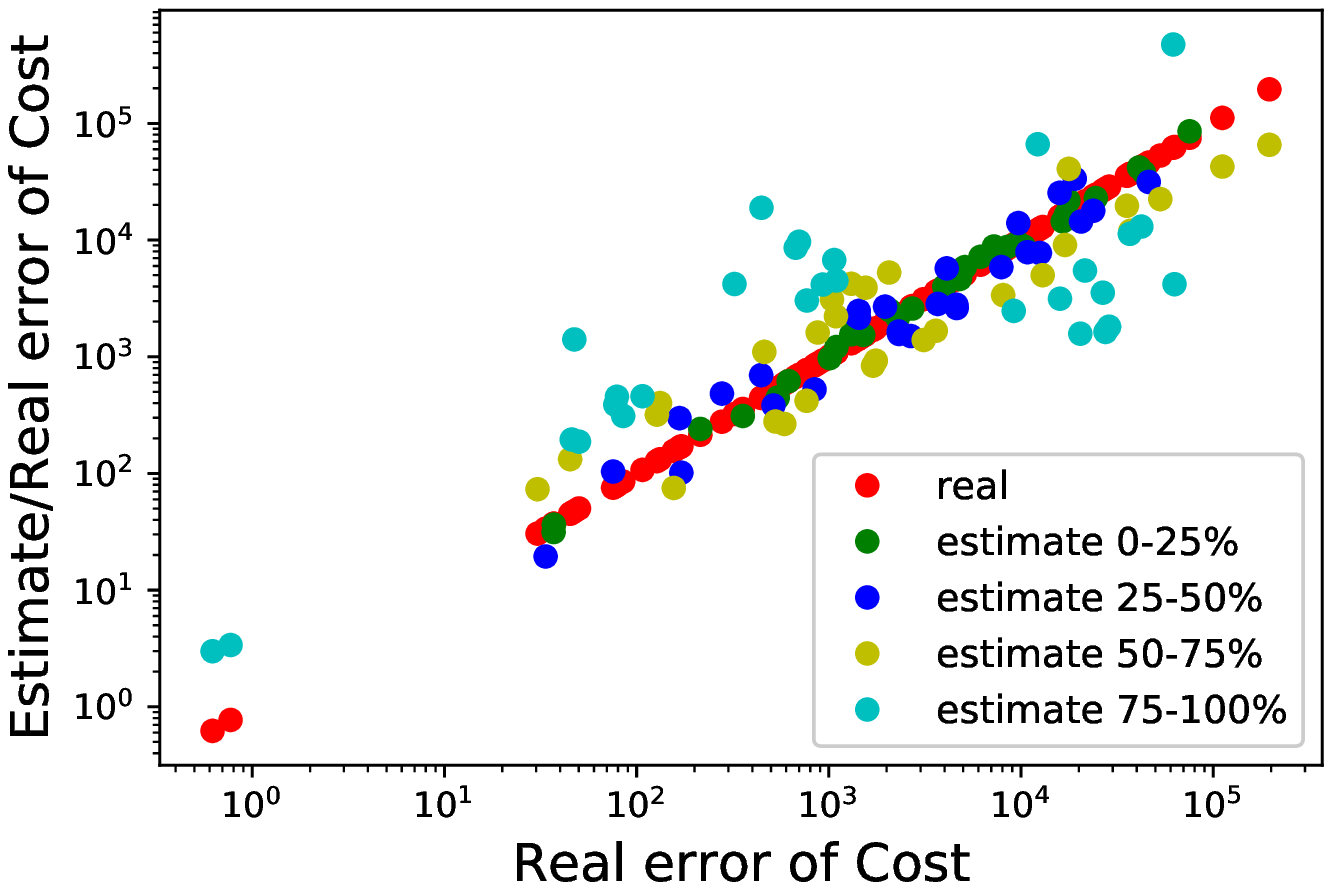}
        \caption{\tpembtcost}
        \label{subfig:search_review_scale}
    \end{subfigure}
    \vspace{-1em}
    \caption[]{Distribution of Estimated Cost}
    \label{fig:job:scatter}
    \vspace{-1em}
\end{figure*}

\subsubsection{Evaluation on the JOB workload}


We train the representation and output layers on 100,000 queries with multiple joins. We take 90\% of multi-table join queries as training data and 10\% of them as validation data. We train the model until the validation cardinality error will not decrease anymore, and then we evaluate the trained model on JOB queries. The results for cardinality estimation are shown in Table~\ref{table:card:job} and the results for cost estimation are shown in Table~\ref{table:cost:job}.

\hi {Hash Bitmap vs String Embedding.}  \tlstmembwcard outperforms \tlstmhashcard on 90-99th errors, max error and mean error for cardinality estimation, because with highly representative of string embedding, the string embedding not only learns the correlations among columns but learns the correlations among tables, and thus the methods with string embedding perform well for cardinality estimation on complex queries with multiple joins. Moreover, the performance difference  between \tlstmembwcost and \tlstmhashcost is even larger. \tlstmembwcost outperforms \tlstmhashcost by 2 times on mean and 99th errors,  and 3 times on max, 95th and 90th errors. Similar to cardinality estimation, the  improvement on complex queries in \tlstmembwcost is due to the correlations carried by string embedding. 

\hi {Rule vs No-Rule.}   \tlstmembtcard outperforms \tlstmembwcard on all the errors for cardinality estimation, especially on 90-99th errors, max error and mean error. The methods with rules achieve better performance, because the rules extract lots of substrings from datasets so that all the strings in the workload are trained and the distribution representations of strings contain more coexistence relations. Complex queries tend to gain more benefit from rules,  because errors would accumulate rapidly for complex queries. Similar to cardinality estimation, \tlstmembtcost outperforms \tlstmembwcost on 90-95th,max and mean errors.

\hi {Tree-Pooling Predicate vs Tree-LSTM Predicate.} The difference between \tlstmembtcard and \tpembtcard is the structure of the predicate embedding model, and \tpembtcard outperforms \tlstmembtcard on all the cardinality errors and \tpembtcost outperforms \tlstmembtcost on all the cost errors. This is because the tree model using the Min-Max Pooling can represent the compound predicate better and train a more robust model for cardinality and cost estimation. On 99th and max errors for cost estimation, \tpembtcost outperforms \tlstmembtcost by 1.5-2 times, because the predicates representation trained by the Min-Max Pooling structure still keeps accurate for complex queries.

\hi{Distribution of errors. } Figure~\ref{fig:job:range} shows the error variance on the JOB workload. For PostgreSQL, we have tuned the factor of page IO so that the unit of the estimated cost equals to the unit of time (milliseconds here). However, it still overestimates  the cost and underestimates the cardinality, and the maximal error is very large in both negative and positive sides.  Our methods underestimate both cost and cardinality of queries, but errors of our methods are  more concentrated and smaller. We adjust the unit of estimated cost to be the same as real execution time by using a factor for each method, and we draw the real time and estimated cost together in Figure~\ref{fig:job:scatter}, which is more intuitive and comprehensive. The cost estimated by \tpembtcost fits the real time very well. \pgcost can not estimate the small cost at all and the estimated cost is very scattered. The error distribution of \tlstmhashcost is between the \tpembtcost and \pgcost.

In summary, the string embedding, using rules to embed the strings, and tree-pooling can improve the quality of cost and cardinality estimation. 

\vspace{-.25em}
\subsection{Efficiency}
\vspace{-.25em}

Table~\ref{table:method:efficient} shows the efficiency of different  methods on 113 queries on the JOB workload. The {\it Batch} indicates whether the batch technique described in Section~\ref{sec:model:train} is used.  The cost evaluation time of {PostgreSQL} is obtained from the planning time without query optimization. \tlstme and \tpe estimate these queries one by one, and \tlstmbe and \tpbe estimate these queries in batch.

\begin{table}[!t]
\setlength{\tabcolsep}{3pt}
\small
\caption{Efficiency Evaluation\label{table:method:efficient}}
\vspace{-1em}
\begin{tabular}{|c|c|c|c|c||c|}
\hline
Methods & Represent & Predicate & Estimate & Batch & Time(ms)\\ \hline
\pg  & No & No & No & No & 18.9 \\ \hline
\mscn  & No & No & SING & No & 14.2\\ \hline
\mscnb  & No & No & SING & Yes & 6.28\\ \hline
\tlstme  & LSTM & LSTM & MULT & No & 70.3\\ \hline
\tlstmbe  & LSTM & LSTM & MULT & Yes & 7.79\\ \hline
\tpe  & LSTM & Pool & MULT & No & 47.3\\ \hline
\tpbe  & LSTM & Pool & MULT & Yes & \textbf{3.63}\\ \hline
\end{tabular}
\vspace{-1em}
\end{table}


\hi{Batch vs No-Batch. } \tpbe outperforms \tpe by one order of magnitude, \tlstmbe also outperforms \tlstme by one order of magnitude. This is because that the batch evaluation reduces the number of times computing the model in nodes of the trees, and increases the parallelism as discussed in Section~\ref{sec:model:train}.

\hi{Tree-Pooling Predicate vs Tree-LSTM Predicate. } \tpe outperforms \tlstme by 50\%, and \tpbe outperforms \tlstmbe by 2 times. This is because tree-pooling replaces some neural networks in the predicate model, and thus has less computation cost in estimation.

%% file: src/sec-con.tex
\vspace{-.5em}
 \section{Conclusion}
\vspace{-.5em}

In this paper, we proposed an end-to-end learning-based tree-structured cost estimator for estimating both cost and cardinality. We encoded query operation, meta data, query predicate and some samples into the model. The model contained embedding layer, representation layer and estimation layer. We proposed an effective method to encode string values into the model to improve the model generalization. Extensive results on real datasets showed that our method outperformed existing techniques.